\documentclass[aps, superscriptaddress, longbibliography,pre,10pt]{revtex4-1}
\usepackage[font=small,labelfont=bf]{caption}
\usepackage[caption=false]{subfig}
\usepackage{amsmath,graphicx,color,epsfig,latexsym,bm,ulem,float,tikz,eucal, mathpazo,times, braket,comment}

\usepackage[colorlinks=true, linkcolor = blue, urlcolor  = blue, citecolor = blue, anchorcolor = red]{hyperref}   

\usepackage{lipsum}
\makeatletter
\newcommand{\shorteq}{%
	\settowidth{\@tempdima}{=}% Width of hyphen
	\resizebox{\@tempdima}{\height}{=}%
}
\makeatletter
\newcommand*\bigcdot{\mathpalette\bigcdot@{.5}}
\newcommand*\bigcdot@[2]{\mathbin{\vcenter{\hbox{\scalebox{#2}{$\m@th#1\bullet$}}}}}
\makeatother
\makeatother
\usetikzlibrary{positioning}
\begin{document}
	%\today
	
	\title{Modified Quantum Wheatstone Bridge based on current circulation}
	\author{Vipul Upadhyay} 
     \email{Corresponding Author}
    
    \email{ vipuupadhyay4@gmail.com }
	\affiliation{Department of Chemistry, Institute of Nanotechnology and Advanced Materials, Center for Quantum Entanglement Science and Technology, Bar-Ilan University, Ramat-Gan, 52900, Israel}
	\author{Rahul Marathe}
    \email{Corresponding Author}
    \email{maratherahul@physics.iitd.ac.in} \affiliation{Department of Physics, Indian Institute of Technology Delhi, Hauz Khas 110 016, New Delhi, INDIA}
	\begin{abstract}{We investigate a simple fermionic system designed to detect an unknown hopping rate between two sites by analyzing current circulation. The system exploits geometric asymmetry and utilizes the connection between the additional energy degeneracy point (AEDP) and current circulation for precise parameter detection. In the low-temperature, low-bias regime, with baths’ chemical potentials aligned near the degenerate energy, a balanced Wheatstone bridge condition emerges when the direction of current circulation reverses, providing a direct means to determine the unknown hopping strength. We further examine the impact of environmental interactions, demonstrating that the device remains functional under moderately strong dephasing and particle losses, though extreme environmental effects eventually degrade performance. Extending the analysis to general operating conditions, it is seen that the device continues to function effectively at higher voltages and temperatures. Finally, an analysis of the quantum Fisher information qualitatively supports our findings, revealing a sharp increase in the coherence contribution and a corresponding decrease in the population contribution near the AEDP. These results highlight geometric asymmetry as a robust and practical tool for quantum metrology.}
	\end{abstract}
	
	\maketitle
	
	\section{Introduction}
    
	Measurement of parameters associated with finite quantum systems can be a challenging task~\cite{Metrology_review_Tth_2014,Quantum_metrology_DeMille2024,quantum_metrology_tomaz2025quantummeasurementproblemreview,metrology_Ullah2025-pl,review_quantum_sensing_RevModPhys.89.035002}. Nonetheless, its importance continues to grow with the ongoing reduction in the size of fundamental components of modern devices.  \textit{Anomalous transport} \cite{anomalous_bijay_PhysRevB.110.L081403,anomalous_PhysRevResearch.7.L012044, Non_trivial_PhysRevE.107.034120,Non_trivial_e21030228,Germometric_CC_1_PhysRevE.99.022131,Geometric_2_PhysRevA.102.023704,Geometry_3_PhysRevE.105.064111} can serve as an ideal candidate in this regard. While the term anomalous transport encompasses several distinct phenomena, in this study we focus on current responses that exhibit nontrivial behavior in the vicinity of specific parameter points of the system \cite{Non_trivial_PhysRevE.107.034120,Non_trivial_e21030228,Germometric_CC_1_PhysRevE.99.022131,Geometric_2_PhysRevA.102.023704,Geometry_3_PhysRevE.105.064111}. Identifying such points can be exploited for precise parameter estimation. 

Numerous efforts have been recently made in this direction. For example various theoretical studies on the Quantum Wheatstone bridge~\cite{Quantum_Wheatstone_Bridge,QWB_New_5x8m-bhgd,QWB_Zhou_2025,QWB_XIE2023106575}, which aim to harness transport anomalies for metrological applications. However, a standardized recipe for constructing such setups remains absent. Addressing this gap forms the central motivation of our present work.  

In recent studies~\cite{Germometric_CC_1_PhysRevE.99.022131,Geometric_2_PhysRevA.102.023704,Geometry_3_PhysRevE.105.064111,Gassab_PhysRevA.109.012424,Upadhyay_2024,Non_trivial_PhysRevE.107.034120} the role of geometric asymmetry in observing anomalous transport was discussed. Specifically in our previous study \cite{Upadhyay_2024}, it was suggested that in quantum systems with some form of asymmetry, circulating currents (CC) can arise in the vicinity of avoided crossings between otherwise non-degenerate energy levels. This idea points toward a general strategy: geometrical asymmetry can act as a control knob to induce CCs, which in turn may serve as a resource for parameter estimation in finite quantum systems. This connection, between geometry, transport anomalies, and metrological utility provides the conceptual foundation of the present study.  

The key aims and novelty of our work are summarized as follows. First, we exploit the role of \textit{geometric asymmetry} in system design, a feature essential for generating circulating currents, but largely overlooked in previous metrology studies. By coupling the baths asymmetrically,  CC gets induced in the system, which forms the basis of our detection scheme. Second, we investigate the \textit{impact of environmental interactions} { as well as random fluctuations in fixed parameters of the system}. These effects are unavoidable in realistic settings, and hence accounting for them is crucial, and it is demonstrated that moderately strong environmental interactions can be tolerated without loss of functionality, provided the interactions do not significantly distort the system’s spectrum. Third, methodologically, the analysis  goes beyond the master equation framework which is valid only in the weak-coupling limit~\cite{breuer2002,Quantum_Wheatstone_Bridge}, and instead the  \textit{Non-Equilibrium Green Function (NEGF)} formalism ~\cite{Bijay1PhysRevB.108.L161115,abhishek_dhar_negf_PhysRevB.73.085119,probe_dvira_10.1063/1.4926395} is employed, which provides an exact treatment for quadratic, non-interacting Hamiltonians~\cite{abhishek_dhar_negf_PhysRevB.73.085119,datta_quantum_transport}.  Within this framework,  analytical results are obtained in the low-bias, low-temperature regime, confirming that the device remains operational even under arbitrarily strong system--bath coupling. Finally, we introduce a \textit{quantitative efficiency metric} to evaluate the device’s performance as a practical measurement tool. { This analysis indicates that combining AEDPs with geometric asymmetry constitutes a stable and practically viable paradigm for sensitive parameter detection, even beyond idealized conditions.} 
\par The manuscript is organized as follows. In Sec.~\ref{model_negf_section}, the model Hamiltonian is introduced together with the NEGF framework. The analysis begins in Sec.~\ref{ideal_sec}, where exact analytical results in the low-bias, low-temperature regime are presented. In Sec.~\ref{environment_sect}, we investigate the influence of environmental interactions using both dephasing and voltage B\"uttiker probes. Sec.~\ref{finite_volt_temp_sec} extends the discussion to finite temperature and voltage, with particular emphasis on the efficiency of the device. Finally, Sec.~\ref{conclusion_sec} summarizes our findings and outlines future research directions.

%%%%%%%%%%%%%%%%%%%%%%%%%%%%%%%%%%%%%%%%%%%%%%%%%%%%%%%%%%%%%%%%%%%%%%%%%%%%%%%%%%%%%%%%%%%%%%%%%%%%%%%%%%%%%%%%%%%%%%%%%%%%%%%%%%%%%%%%%%%%%%%%%%%%%%%%%%%%%%%%%%%%%%%%%%%%%%%%%%%%%%%%%%%%%%%%%%%%%%%%%%%%%%%%%%%%%%%%%%%%
	\section{Model and Methodology} \label{model_negf_section}
\begin{figure}[t]
		\centering 
		{\includegraphics[width=0.3\linewidth,height=0.3\linewidth]{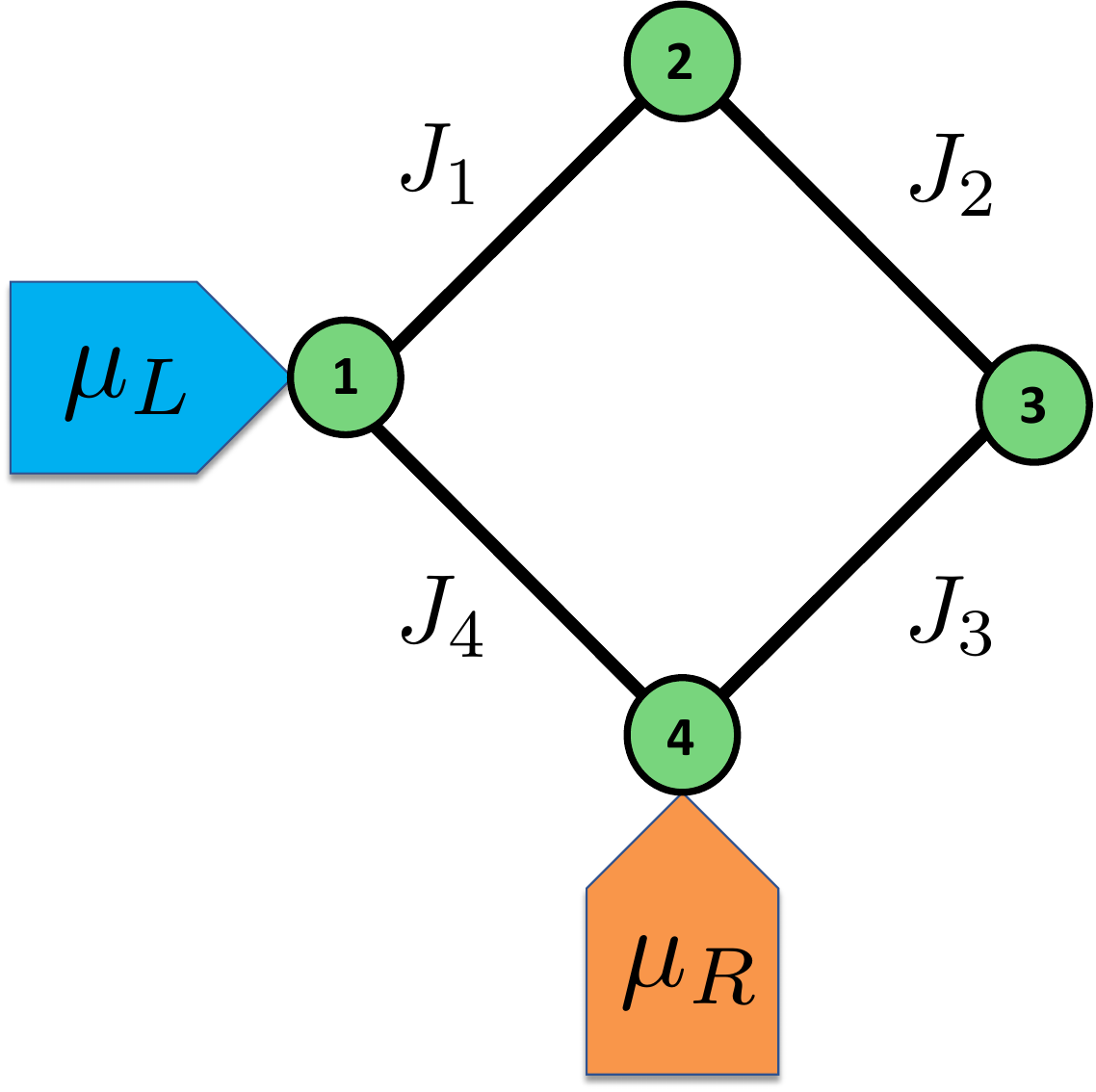}}
		\caption{Schematic diagram of the proposed model. The modified Wheatstone bridge is asymmetrically connected to the baths, such that the upper branch $(1\to2\to3\to4)$ has more number of fermionic sites than the lower branch ($1\to4$) between the baths. Also, the usual link between sites 2 and 4 is not present. The hopping rate $\textnormal{J}_1$ and $\textnormal{J}_4$ are fixed \cite{Quantum_Wheatstone_Bridge}. We want to determine the hopping rates $\textnormal{J}_2$ by slowly varying the controllable parameter $\textnormal{J}_3$. $\textnormal{J}_2$ can be determined by using the balanced Wheatstone bridge condition, $\textnormal{J}_2=\frac{\textnormal{J}_1}{\textnormal{J}_4} \textnormal{J}^{0}_3$, where $\textnormal{J}^{0}_3$ is the point where the current circulation reverses direction.}
		\label{Model}
	\end{figure}

	We begin by introducing the model system (see Fig. \ref{Model}). This model is slighty different from the usual Wheatstone bridge \cite{Quantum_Wheatstone_Bridge} as it does not have the connecting link between sites numbered $2$ and $4$.
    This is the simplest model which allows us to see the balanced Wheatstone bridge condition. Also, as  will be seen later, the single particle energy spectrum of this system has an additional energy degeneracy point (AEDP) when the Wheatstone bridge condition is satisfied. This feature makes it a good model for studying the utility of the anomalous transport in designing metrology devices. The Hamiltonian for the system is given as,
	\begin{align} \label{Hamiltonian}
		\hat{H}_0=\mu\sum_{n=1}^4 \hat{c}_n^\dagger \hat{c}_n+\sum_{n=1}^4\textnormal{J}_n(\hat{c}^\dagger_n\hat{c}_{n+1}+\hat{c}^\dagger_{n+1}\hat{c}_{n}).
	\end{align}
	where, $\hat{c}_n^{ \dagger} (\hat{c}_n)$ is the creation and destruction operator at the site $n$,  $\mu$ is the onsite chemical potential and $\textnormal{J}_n$ is the hopping rate from site $n$ to $n+1$ (n=5 becomes n=1 due to periodic boundary condition).  This modified Wheatstone bridge is connected to two fermionic thermal baths at different chemical potentials $\mu_L$ and $\mu_R$ respectively.  The right bath is connected at site numbered `4' and the  left bath at site numbered `1'. This results in the upper branch having more fermionic sites than the lower branch, which forms the source of asymmetry in our system. The implications of combining this geometry asymmetry with AEDP will be later discussed in this manuscript \cite{Germometric_CC_1_PhysRevE.99.022131, Upadhyay_2024}. The system-bath interaction is given by the standard hopping type Hamiltonian,
	\begin{align}\label{SB}
		\hat{H}_{SB}=&\hat{c}_1^{\dagger}\otimes  \sum_k g^L_k  \hat{b}^L_k + \hat{c}_1\otimes  \sum_k g^{L*}_k  \hat{b}_k^{L\dagger} +\hat{c}_{4}^{\dagger}\otimes  \sum_k g^R_k  \hat{b}^R_k + \hat{c}_{4}\otimes  \sum_k g^{R*}_k  \hat{b}_k^{R\dagger} 
	\end{align}
	where $g^L_k ( g^R_k)$ denote the system-bath interaction strength for left (right) terms. Finally, the Hamiltonian for the Fermionic baths is given as,
	\begin{align}\label{Bath hamiltonian}
		\hat{H}_B=\sum_n \omega^L_n \hat{b}_n^{L \dagger}\hat{b}^L_n+\sum_{n}\omega^R_n \hat{b}_n^{R \dagger}\hat{b}^R_n
	\end{align}	
	where, $\hat{b}_n^{i \dagger} (\hat{b}_n^{i})$ is the creation (destruction) operator of the  $n^{th}$ mode of the $i^{th}$ bath. Our aim is  to detect the \textit{unknown parameter $\textnormal{J}_2$}. This is achieved by varying the controllable parameter $\textnormal{J}_3$ with the other hopping parameters, $\textnormal{J}_1$ and $\textnormal{J}_4$, kept constant. Experimental approaches for realizing such a Hamiltonian can be found in \cite{exp_same_2,exp_same_PhysRevX.7.031001} and in the supplementary material of \cite{Quantum_Wheatstone_Bridge}.
The single-particle energy eigenvalues of the model are given by,

%%%%%%%%%%%%%%%%%%%%%%%%%%%%%%%%%%%%%%%%%%%%%%%%%%%%%%%%%%%%%%%%%%%%%%%%%%%%%%%%%%%%%%%%%%%%%%%%%%%%%%%%%%%%%%%%%%%%%%%%%%%%%%%%%%%%%%%%%%%%%%%%%%%%%%%%%%%%%%%%%%%%%%%%%%%%%%%%%%%%%%%%%%%%%%%%%%%%%%%%%%%%%%%%%%%%%%%%%%%%
    
\begin{equation}
	E^{3,4}_{1,2} = \mu \pm \sqrt{\frac{E_0^2 \pm \sqrt{E_0^4 - 4 \xi^2}}{2}},
\end{equation}  
where $E_0^2 = \textnormal{J}_1^2 + \textnormal{J}_2^2 + \textnormal{J}_3^2 + \textnormal{J}_4^2$  and	$\xi \equiv \textnormal{J}_1 \textnormal{J}_3 - \textnormal{J}_2 \textnormal{J}_4$. It follows that when $\xi = 0$, one obtains the condition  
\begin{align} \label{Wheatstone}
	\frac{\textnormal{J}_1}{\textnormal{J}_4} = \frac{\textnormal{J}_2}{\textnormal{J}_3}
\end{align}  
which is identical to the standard Wheatstone bridge balance condition. 
When this condition is satisfied, the system exhibits an AEDP in its spectrum (see Fig.~\ref{ideal_fig} (a)). The emergence of  AEDP can be understood as a consequence of the enhanced symmetry in the system introduced by the Wheatstone balance. 
This additional symmetry collapses two otherwise distinct energy levels, resulting in additional spectrum degeneracy.  

Importantly, the AEDP has significant transport implications. 
According to the analysis in Ref.~\cite{Upadhyay_2024}, introducing a small asymmetry in systems with AEDP can give rise to anomalous current behavior in the vicinity of the AEDP. Such currents have a Fano resonance \cite{Fano_ref} like anomalous behavior. 
This behavior is characterized by sharp conductance variations and non-trivial current asymmetries, which are absent away from the degeneracy. 
{\par From an experimental perspective, a variety of platforms can be utilized to study the model discussed in this manuscript. These include nanoscale electronic circuits~\cite{expp2_Jia2019,expp3_Ronzani2018}, cold-atom systems~\cite{exp_cold_atoms,exp_cold_atoms1}, trapped ions~\cite{exp_trapped_ion,exp_trapped_ion2}, molecular junctions~\cite{Quantum_metrology_DeMille2024}, NMR setups \cite{exp1_Micadei2019}, and quantum dots~\cite{expp4_Hendrickx2021,exp_quantum_dots,exp_quantum_dot_2,typical_values10.1063/1.5025928}. 
Each of these paradigms provides the essential capabilities required for implementing the proposed scheme, namely tunable system--bath couplings, precise control over geometric configurations, and the ability to measure current correlations. Such capabilities are already available in state-of-the-art experimental setups.
\par { Additionally, all parameters are expressed relative to the typical hopping strength, the actual value of which depends on the details of the specific experimental setup \cite{exp_same_PhysRevX.7.031001,exp_same_2}.  For example, if the model is designed using a Quantum dot setup similar to the study \cite{typical_values10.1063/1.5025928}, typical hopping strengths are of the order $10 \mu eV$. In temperature units this translates to `$1 \sim 0.1$K' in the mentioned units, suggesting that the bridge works optimally at low temperatures. However, if some other physical setup is considered where the hopping strength is much larger, the device should also function efficiently for much larger temperatures. Similarly, the voltage is measured in same scale as the hopping strength. Finally, the particle current has dimensions of inverse time, with the characteristic timescale determined again by the inverse of hopping strength. }
}

    \subsection{Non-Equilibrium Green Function (NEGF) Equations} \label{appendix_negf}
    Since the Hamiltonian given in eq. \eqref{Hamiltonian} is a non-interacting one, the standard NEGF equations are utilized for analyzing the model. Such an analysis  helps us to understand the behavior of the device for arbitrarily strong system-bath coupling, which cannot be done using the master equation approach \cite{breuer2002,datta_quantum_transport}. The equations for this method are given below.
    \par First, due to the coherent transport setup, the total particle current is given by the Landauer-B\"{u}ttiker formula \cite{datta_quantum_transport},
	\begin{align}
		I_T=\frac{1}{2\pi}\int^{\infty}_{-\infty} dE~ T(E)~ (f_L(E)-f_R(E))
	\end{align}
	where, we work in the natural units $\hbar=1,k_B=1,e=1$ throught the manuscript.  The transmission is given as	$T(E)=Tr[\Gamma_1G^R\Gamma_4G^A]$, which depends on the retarded Green Function given as, \cite{abhishek_dhar_negf_PhysRevB.73.085119,datta_quantum_transport} ,
	\begin{align}
		G^R\equiv G^R(E)=\big{(}E-H+\frac{i}{2}( \Gamma_1+ \Gamma_4)\big{)}^{-1}
	\end{align}
	and, $G^A\equiv G^A(E)=(G^R)^\dagger$. Also, $H$ is the matrix corresponding to the Hamiltonian in eq. \eqref{Hamiltonian}, such that $\hat{H}_0=\sum_{nm}H_{nm}\hat{c}_n^\dagger \hat{c}_m$, and   the matrices $(\Gamma_{1(4)})_{nm}=\gamma \delta_{nm} \delta_{n1(4)}$, with  $\gamma$ being related to the square of system bath interaction strengths $ |g^{L(R)}_k|^2$.  We take the wide band limit \cite{probe_dvira_10.1063/1.4926395,wide_band_ridley_PhysRevB.95.165440} such that $\gamma$ is not a function of energy. To calculate the branch currents,  the following expression for local current at site `n' \cite{Prosen_2008_NJP} is utilized,
	\begin{align}
		I_n =2\textnormal{J}_n Im(\hat{C}_{n,n+1}).
	\end{align}
	where, the single particle correlation matrix \cite{datta_quantum_transport} is defined as,
	\begin{align}
		\hat{C}_{nm}\equiv\langle \hat{c}^\dagger_m \hat{c}_n \rangle
	\end{align}
where, the index notations are reversed to make the relations with Green functions more convenient. The correlation matrix  is related to the energy resolved average occupancy green function $G^N(E)=-i G^<(E)$ \cite{datta_quantum_transport},
	\begin{align}
		\hat{C}_{nm}=\frac{1}{2\pi}\int^{\infty}_{-\infty} dE G_{nm}^N(E)
	\end{align}
	where, $G^N(E)=f_L G^R\Gamma_1G^A+f_R G^R\Gamma_4G^A$ \cite{probe_dvira_10.1063/1.4926395}, for the system-bath coupling parameter $\gamma$. Doing some simplifications, it is found 
    that the upper $(I_U)$ and lower branch current $(I_D)$ can be evaluated by finding the local current from site $1\to2$ and $1\to4$ and are given as, 
	\begin{align} \label{eq_u_cuurent}
		I_U&=2\textnormal{J}_1 Im(\hat{C}_{12})=\frac{1}{2\pi}\int^{\infty}_{-\infty} dE ~T_U(E)~ (f_L(E)-f_R(E)) \nonumber \\
		I_D&=2\textnormal{J}_4 Im(\hat{C}_{14})=\frac{1}{2\pi}\int^{\infty}_{-\infty} dE~ T_D(E)~ (f_L(E)-f_R(E))
	\end{align}
	with, $ T_U(E) = 2\gamma \textnormal{J}_1 ~Im(G^R_{11}G^A_{12}), ~T_D(E) = 2\gamma \textnormal{J}_4~ Im(G^R_{11}G^A_{14})$. Finally, the Fermi-Dirac distribution function for a lead at chemical potential $\mu_i$ is defined as,
    \begin{align}
        f_i(E)=\frac{1}{1+e^{\frac{(E-\mu_i)}{T}}}
    \end{align}
    In what follows, we use the above quantities to analyze the behavior of the currents in different parameter regimes. Additionally, according to our convention, all the currents are positive when the current flows from left to right. So, CC occurs when one of the branch current has negative sign.
  
%%%%%%%%%%%%%%%%%%%%%%%%%%%%%%%%%%%%%%%%%%%%%%%%%%%%%%%%%%%%%%%%%%%%%%%%%%%%%%%%%%%%%%%%%%%%%%%%%%%%%%%%%%%%%%%%%%%%%%%%%%%%%%%%%%%%%%%%%%%%%%%%%%%%%%%%%%%%%%%%%%%%%%%%%%%%%%%%%%%%%%%%%%%%%%%%%%%%%%%%%%%%%%%%%%%%%%%%%%%%
	
	\section{Conductance in the low-temperature, low-bias regime} \label{ideal_sec}
   
	\par We begin our analysis by first focusing on the low temperature, low voltage regime. This is valid if the system is at sufficiently low temperature such that the ratio of the system’s characteristic energy scale to the temperature tends to infinity. This coupled with very small voltage bias allows us to get rid of  the energy  integrals in section \ref{appendix_negf} \cite{TUR_BijayPhysRevB.98.155438}, and the currents can be expressed directly in terms of conductance at the on site chemical potential $\mu$.
	\begin{align}
		&I_T= \frac{1}{2\pi}T(\mu) V,
		&&I_U= \frac{1}{2\pi}T_U(\mu) V,
		&&&I_D=\frac{1}{2\pi} T_D(\mu) V.
	\end{align}
    Here, the chemical potentials of the two leads are $\mu_{L(R)}=\mu \pm V/2$, $T(\mu), T_U(\mu)$ and $T_D(\mu)$ are the transmissions for the three currents which are equal to the $2 \pi$ times the conductance in this regime \cite{datta_quantum_transport,Bijay1PhysRevB.108.L161115}. It is possible to analytically calculate the retarded Green function $G^R (\mu)$ at the energy $\mu$ explicitly for the given model.  The expression for the conductance for the three currents defined in \ref{appendix_negf} are given as,
\begin{align} \label{eqs_conductance}
		&\mathcal{G}_U=\frac{1}{\pi}\frac{8\gamma^2\xi \textnormal{J}_2^2 \textnormal{J}_1 \textnormal{J}_3}{(4 \xi^2+\gamma^2 \textnormal{J}_2^2)^2} ,
		&&\mathcal{G}_D=-\frac{1}{\pi}\frac{8\gamma^2\xi \textnormal{J}_2^3 \textnormal{J}_4}{(4 \xi^2+\gamma^2 \textnormal{J}_2^2)^2},
		&&&\mathcal{G}_T=\frac{1}{\pi}\frac{8\gamma^2\xi^2 \textnormal{J}_2^2 }{(4 \xi^2+\gamma^2 \textnormal{J}_2^2)^2}.
	\end{align}
  \begin{figure}[t]
		\centering 
		\subfloat []
		{\includegraphics[width=0.32\linewidth,height=0.27\linewidth]{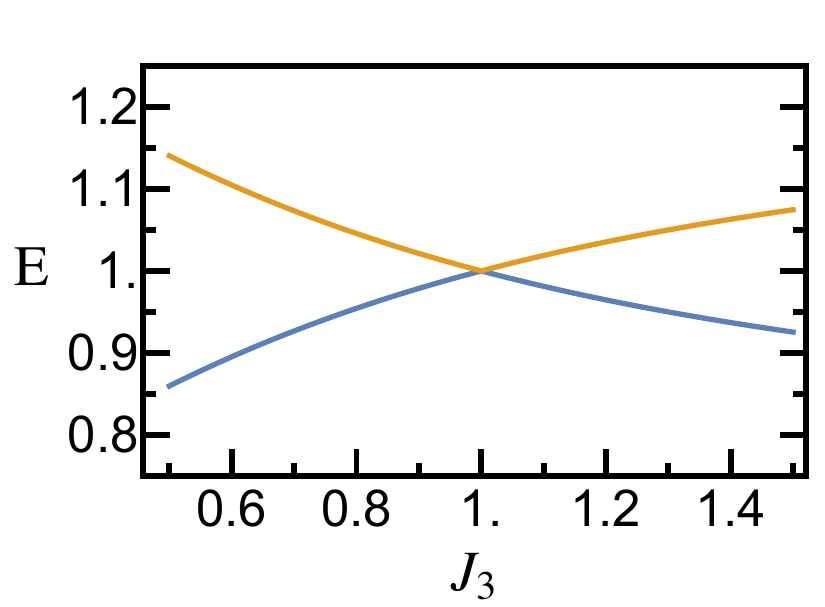}}
		\subfloat []
		{\includegraphics[width=0.32\linewidth,height=0.27\linewidth]{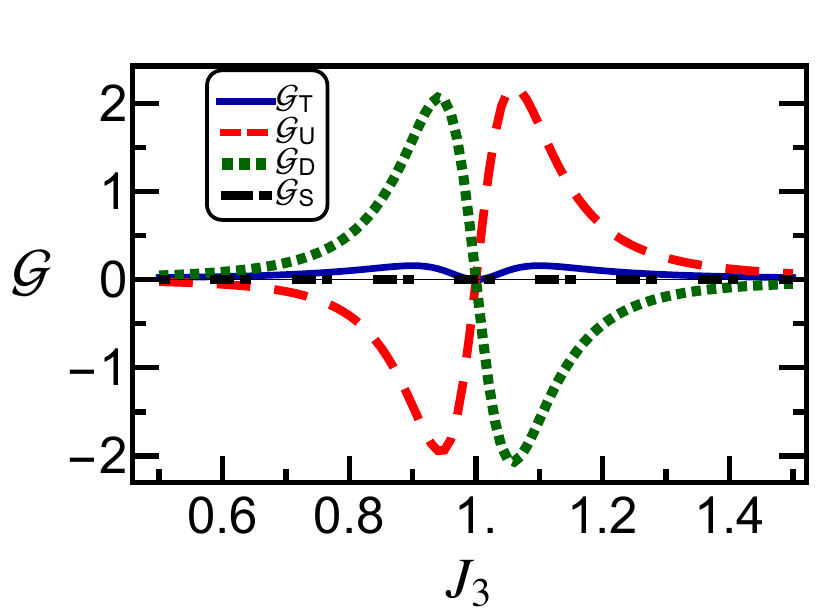}}
	\subfloat []
        {\includegraphics[width=0.32\linewidth,height=0.27\linewidth]{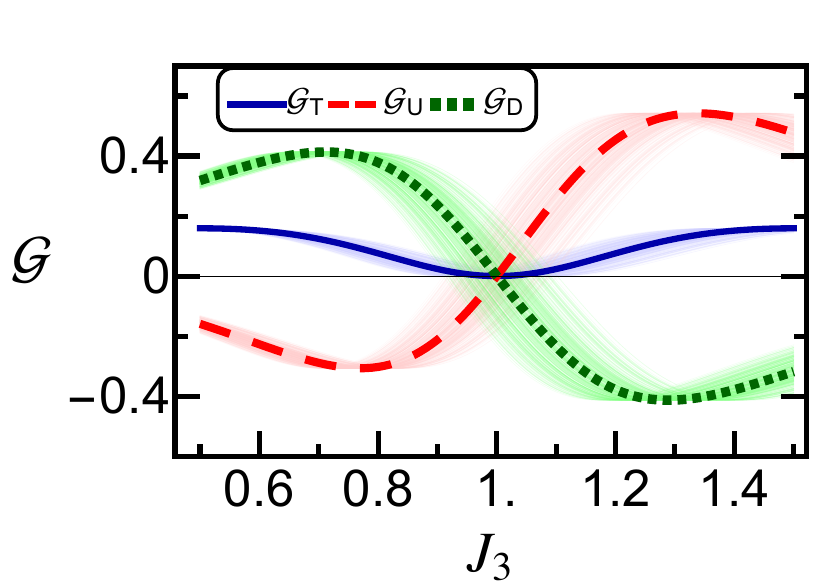}}
		\caption{\textbf{(a)} Location of additional degeneracy in the energy spectrum of the single particle Hamiltonian. Variation of total $(\mathcal{G}_T)$, upper $(\mathcal{G}_U)$ and lower $(\mathcal{G}_D)$ branch conductance with the controllable hopping parameter $\textnormal{J}_3$ for \textbf{(b)} Weak system bath coupling with $\gamma=0.1$, $(\mathcal{G}_S)$ denotes the upper branch conductance when the baths are symmetrically connected to the model. \textbf{(c)} Strong system bath coupling with $\gamma=0.5$, { the color bands show the results for $5 \%$ uncertainty in the fixed parameters $\textnormal{J}_1,\textnormal{J}_4$ for 500 different realisations}. For all the figures, the default value of parameters if not specifically  defined are $\mu=1,\textnormal{J}_1=0.25, \textnormal{J}_4=0.5,\gamma=0.5$. The hopping rate to be determined is set at $\textnormal{J}_2=0.5$. All parameters are expressed relative to the typical hopping strength, the actual value of which depends on the details of the specific experimental setup \cite{exp_same_PhysRevX.7.031001,exp_same_2}. { For example, if the model is designed using a Quantum dot setup similar to the study \cite{typical_values10.1063/1.5025928}, typical hopping strengths are of the order $10 \mu eV$. In temperature units this translates to `$1 \sim 0.1$K' in the mentioned units, suggesting that the bridge works optimally at low temperatures. However, if some other physical setup is considered where the hopping strength is much larger, the device should also function efficiently for much larger temperatures. Similarly, the voltage is measured in same scale as the hopping strength. Finally, the particle current has dimensions of inverse time, with the characteristic timescale determined again by the inverse of hopping strength. } }
		\label{ideal_fig}
	\end{figure}
where $\mathcal{G}_u, \mathcal{G}_D$ and $\mathcal{G}_T$ are the up, down and total conductances respectively.
Looking at the analytical expressions above, it is observed that the conductances corresponding to the upper and lower branch currents change sign whenever the parameter $\xi$ changes sign, while the conductance of the total current remains strictly positive at all times. Furthermore, the conductances of the two branches always possess opposite signs for the entire range of parameter values considered. These conductances also satisfy Kirchhoff’s conservation rule, namely $\mathcal{G}_T=\mathcal{G}_U+\mathcal{G}_D$. A closer examination of the expression reveals that the magnitude of the total current is always less than or equal to the magnitude of one of the branch currents in the bridge configuration, which necessarily implies the presence of circulating particle currents within the branches themselves. Importantly, the sense of this circulation reverses, from clockwise to anticlockwise, precisely at the point where $\xi$ crosses zero and becomes positive. This reversal forms the \textbf{key idea behind our device}: by carefully tracking the direction of the branch currents and pinpointing the parameter value at which their sign changes, one can deduce the value of the unknown parameter $\textnormal{J}_2$. This determination can be achieved using the Wheatstone condition given in equation \eqref{Wheatstone}. In particular, if the controllable parameter $\textnormal{J}_3$ takes the value $\textnormal{J}_3^0$ at the moment the reversal occurs, then the unknown parameter follows directly as,
\begin{align}
\textnormal{J}_2 = \frac{\textnormal{J}_1}{\textnormal{J}_4}  \textnormal{J}_3^{0}.
\end{align}
It is also observed that the mathematical expressions remain valid for all orders of the system–bath coupling parameter $\gamma$, indicating that the connection between AEDP and current circulation is considerably stronger than what was proposed in the earlier study \cite{Upadhyay_2024}, where this relationship was established only within the regime of weak system–bath coupling. Having now discussed the analytical results in some detail, we turn to the numerical observations and examine the behavior of the current as a function of the variable hopping strength $\textnormal{J}_3$, as illustrated in Fig.~\ref{ideal_fig}.
	\par Our analysis in Fig. \ref{ideal_fig} reveals that the reversal of current circulation (CC) occurs even when the system--bath coupling is strong, demonstrating that this effect is not restricted to the weak-coupling regime. However, when the coupling is weak,  the branch currents attain larger magnitudes in the vicinity of the additional degeneracy point, in agreement with the mathematical form of Eq.~\eqref{eqs_conductance} as can be seen in Fig. \ref{ideal_fig} (b) and (c).   In the special case where the coupling to the baths is perfectly symmetric (at the sites `1' and `3'), the transmission contribution at the energy of interest vanishes uniformly, as illustrated by the black dotted lines in Fig.~\ref{ideal_fig} (b). 
    This indicates that the device only works when there is some asymmetry as discussed in some earlier studies \cite{Upadhyay_2024,Geometric_2_PhysRevA.102.023704,Geometry_3_PhysRevE.105.064111,Germometric_CC_1_PhysRevE.99.022131}. { Additionally, as can be seen in Fig. \ref{ideal_fig} (c), if there is uncertainty in the value of the fixed hopping parameters, $\textnormal{J}_1,\textnormal{J}_4$, the CC still persists, however the point at which CC changes direction shifts. This may bring about some errors in the determination of the unknown parameter $\textnormal{J}_2$. To quantify these errors, we revisit the balanced condition, according to which, the value of the tunable parameter $\textnormal{J}_3$ at which the CC changes sign is given as,
    \begin{align}
\textnormal{J}_3 = \frac{\textnormal{J}_4}{\textnormal{J}_1}  \textnormal{J}^0_2=\frac{\textnormal{J}^0_4\pm\Delta \textnormal{J}_4}{\textnormal{J}^0_1\pm\Delta \textnormal{J}_1}  \textnormal{J}^0_2
\end{align} 
where `$\textnormal{J}^0, \Delta \textnormal{J}$' respectively specify the ideal values and fluctuations of the various hopping parameters $\textnormal{J}_1,\textnormal{J}_4,\textnormal{J}_2$. Based on this $\textnormal{J}_3$, the value determined for $\textnormal{J}_2$ is,
\begin{align}
\textnormal{J}_2 &= \frac{\textnormal{J}^0_1}{\textnormal{J}^0_4}  \textnormal{J}_3=\frac{\textnormal{J}_4}{\textnormal{J}_1}  \textnormal{J}_2=\frac{\textnormal{J}^0_1}{\textnormal{J}^0_4}\frac{\textnormal{J}^0_4\pm\Delta \textnormal{J}_4}{\textnormal{J}^0_1\pm \Delta \textnormal{J}_1}  \textnormal{J}^0_2 \nonumber\\
&=\left(1\pm \frac{\Delta \textnormal{J}_4}{\textnormal{J}_4^0} \mp \frac{\Delta \textnormal{J}_1}{\textnormal{J}_1^0}+\mathcal{O}((\Delta \textnormal{J}/\textnormal{J}^0)^2)\right)\textnormal{J}^0_2
\end{align}
The above expression tells us that the error in determining $\textnormal{J}_2$ grows linearly with relative errors in the fixed parameters, for small uncertainty in their values. This means that the device functions efficiently for small fixed parameter fluctuations.
}.  The above observations collectively indicate that the operational principle of our device is robust across the entire range of system-bath interaction strengths as well {as small fixed parameter fluctuations}.
    However, to fully assess its practical performance, it is crucial to examine in detail how interaction with the  surrounding environment influences its behavior.
%%%%%%%%%%%%%%%%%%%%%%%%%%%%%%%%%%%%%%%%%%%%%%%%%%%%%%%%%%%%%%%%%%%%%%%%%%%%%%%%%%%%%%%%%%%%%%%%%%%%%%%%%%%%%%%%%%%%%%%%%%%%%%%%%%%%%%%%%%%%%%%%%%%%%%%%%%%%%%%%%%%%%%%%%%%%%%%%%%%%%%%%%%%%%%%%%%%%%%%%%%%%%%%%%%%%%%%%%%%%	
    
    \section{Interaction with the environment} \label{environment_sect}
Since current circulation is fundamentally a wave phenomenon \cite{CC_classical_rahul_sir,Upadhyay_2024}, arising from resonances near certain states, it is reasonable to expect that introducing interactions with the environment could diminish this effect. This possibility in our model using two approaches: first, by employing virtual voltage B\"{u}ttiker probes \cite{Bijay_voltage_probes_PhysRevB.105.224204,Bijay1PhysRevB.108.L161115} to induce dephasing while retaining elastic scattering within the bridge, and second, by allowing particle leakage \cite{Bijay_lossy_PhysRevB.110.235425}. Such effects naturally arise when the particles in our system interact with other degrees of freedom, such as electrons or phonons \cite{probe_dvira_10.1063/1.4926395}. The following subsections report results obtained in the low-temperature, low- voltage bias limit. Also, the energy $E$ for all the Green function is set equal to the on site potential $\mu$.

	\subsection{Addition of zero current B\"{u}ttiker Probes}
	We first study the effect of dephasing induced by virtual  B\"{u}ttiker probes \cite{probe_dvira_10.1063/1.4926395} in our system. To implement this,  virtual B\"{u}ttiker broadening leads are connected to all sites of the system, and their chemical potentials are adjusted such that the net particle current in each virtual lead remains zero. This condition is enforced by introducing virtual baths similar to the real ones defined in Eq.~\eqref{Bath hamiltonian} and coupling them to the system through the same form of interaction, Eq.~\eqref{SB},  with coupling strength $\gamma_P$. The inclusion of these probes consequently modifies the retarded Green's function \cite{Bijay1PhysRevB.108.L161115}, and which the following form.

	\begin{align}
		G^R=\big{(}E-H+\frac{i}{2}( \Gamma_1+ \Gamma_4)+\frac{i}{2}\sum_m \Gamma^P_m\big{)}^{-1}.
	\end{align}
Similarly, the energy dependent  correlation matrix contributions get modified as,
\begin{equation}
    G^N = f_L G^R \Gamma_1 G^A + f_R G^R \Gamma_4 G^A + \sum_m f_m G^R \Gamma^P_m G^A ,
\end{equation}  
where the system–probe B\"{u}ttiker Broadening matrix is denoted by $\Gamma^P$ with {$(\Gamma^P_m)_{i,j}=\gamma_P \delta_{ij}\delta_{jm}$}. The chemical potentials $f_m$ of the virtual baths are determined by imposing the constraint that the net particle current in each virtual bath is zero. In this setup, it is possible to explicitly solve for all the chemical potentials, leading to the following modified expressions for the conductances \cite{Bijay1PhysRevB.108.L161115,probe_dvira_10.1063/1.4926395}. Specifically, refer to the supplementary material of the study \cite{Bijay1PhysRevB.108.L161115}.
	\begin{align} \label{Eqs_buttik    er_probe}
		&\mathcal{G}_T=\frac{1}{2\pi}\left(\gamma^2 |G^{R}_{1,4}|^2+\gamma^2\gamma_P\sum_{n,m}|G^{R}_{n,4}|^2W^{-1}_{n,m}|G^{R}_{m,1}|^2 \right),\nonumber \\&\mathcal{G}_U=\frac{\textnormal{J}_1}{\pi} Im(G^N_{12})=\frac{\textnormal{J}_1}{\pi}\left(\gamma Im(G^R_{1,1}G^A_{1,2})+\gamma\gamma^2_P\sum_{mn} Im(G^R_{1,m}G^A_{m,2})W^{-1}_{mn}|G^R_{n1}|^2\right)
\nonumber \\
		&\mathcal{G}_D=\frac{\textnormal{J}_4}{\pi}\left(\gamma Im(G^{R}_{11}G^{A}_{14})+\gamma\gamma^2_P\sum_{mn} Im(G^R_{1m}G^A_{m4})W^{-1}_{mn}|G^R_{n1}|^2\right)
	\end{align}
 
\begin{figure}[t]
		\centering 
		\subfloat []
		{\includegraphics[width=0.32\linewidth,height=0.27\linewidth]{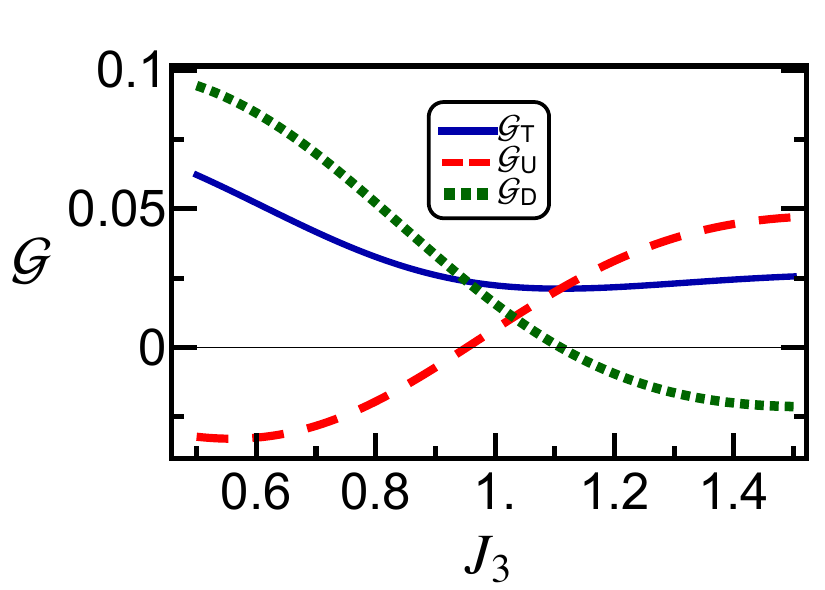}}
		\subfloat []
		{\includegraphics[width=0.32\linewidth,height=0.27\linewidth]{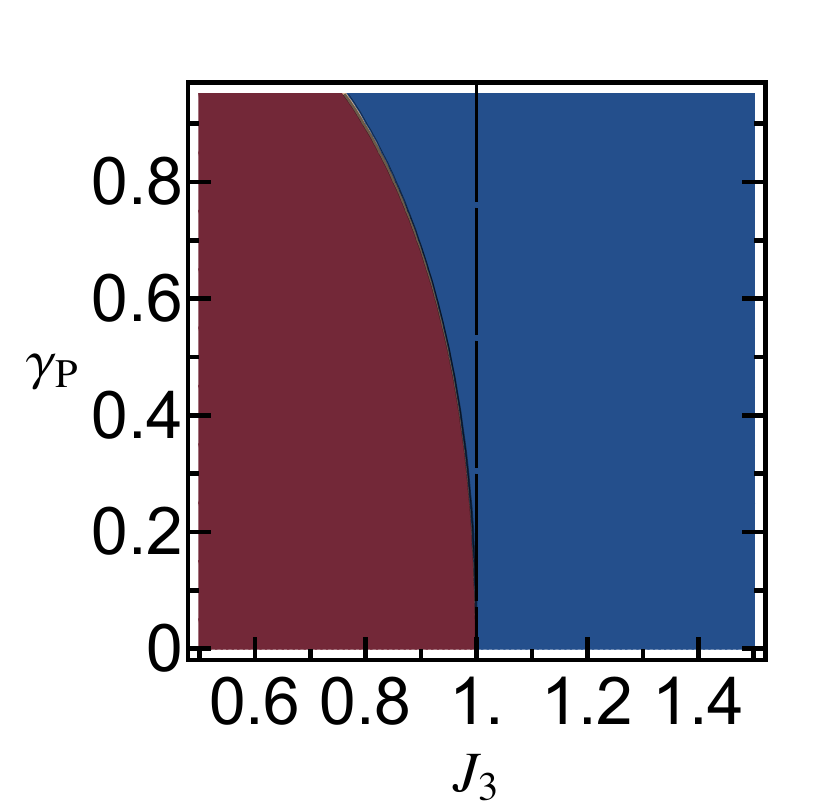}}
		\subfloat []
		{\includegraphics[width=0.32\linewidth,height=0.27\linewidth]{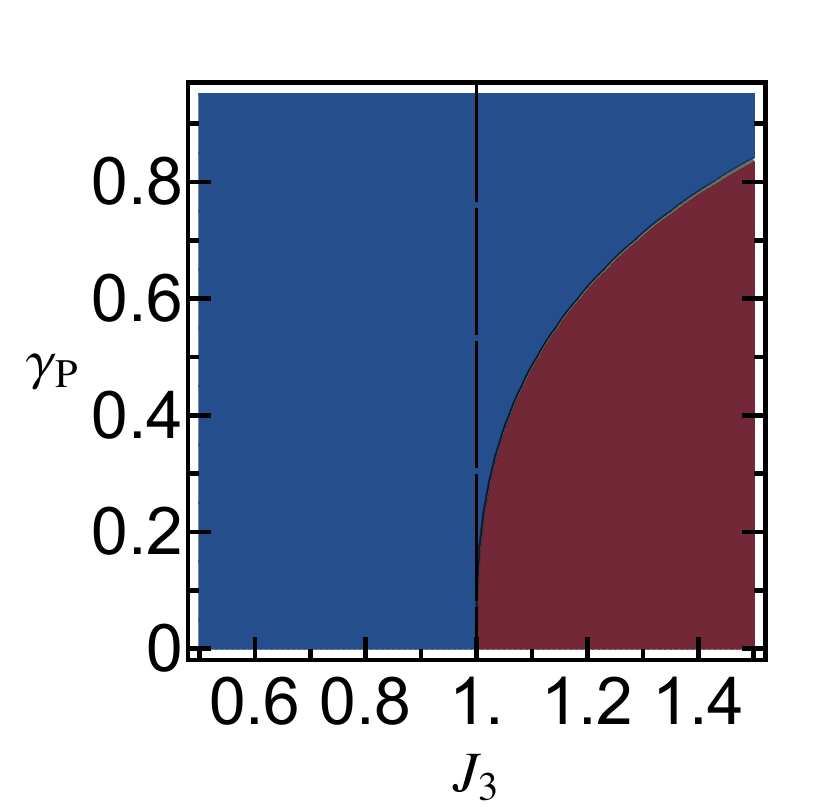}} \\
        \subfloat[]
		{\includegraphics[width=0.32\linewidth,height=0.27\linewidth]{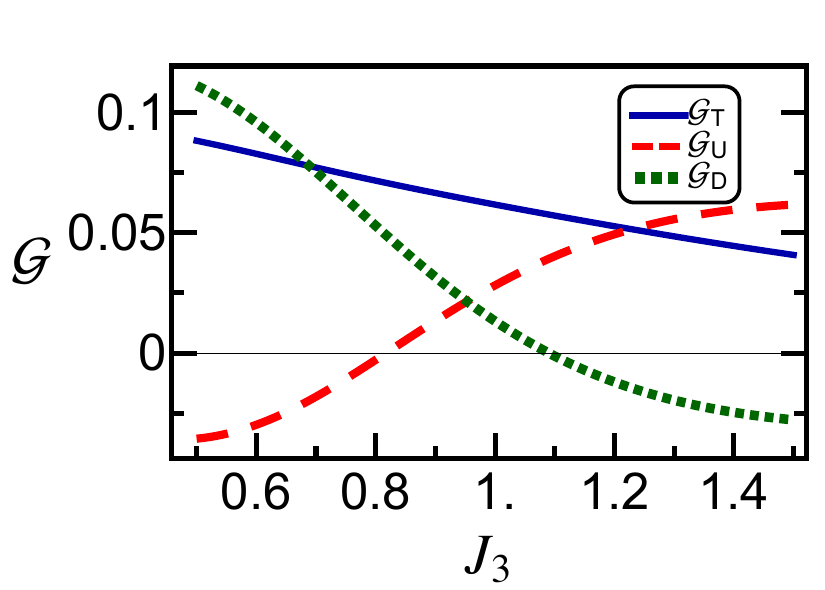}}
        \subfloat[]
		{\includegraphics[width=0.32\linewidth,height=0.27\linewidth]{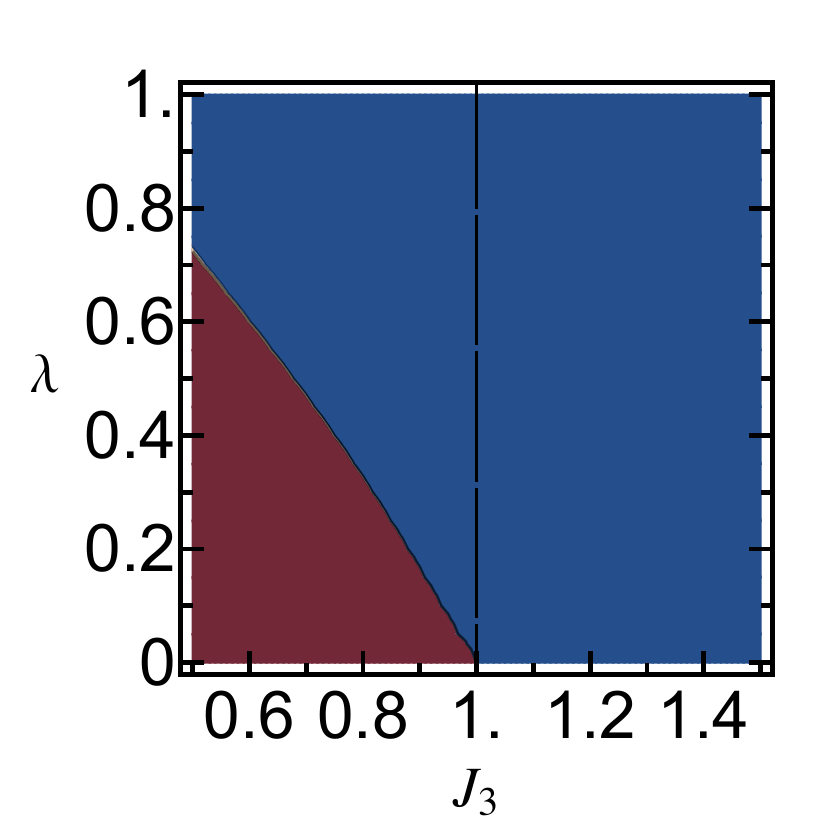}}
		\subfloat[]
		{\includegraphics[width=0.32\linewidth,height=0.27\linewidth]{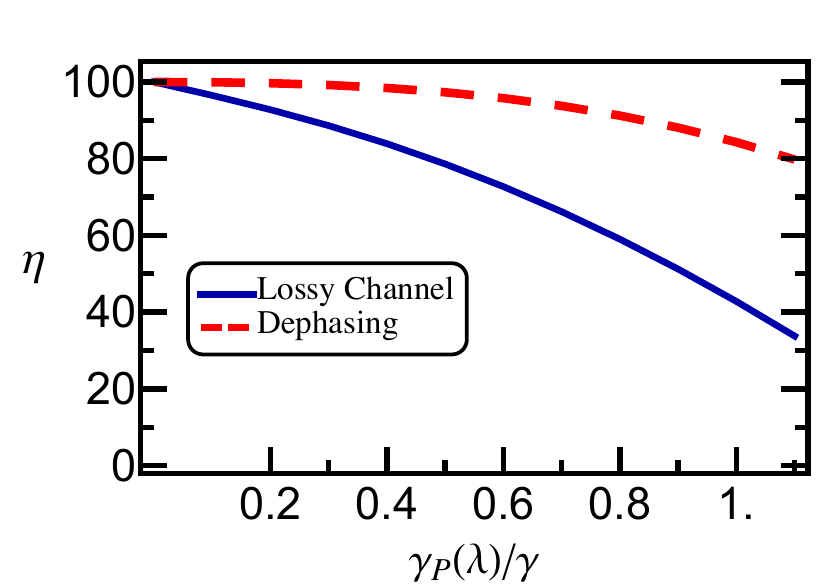}}
		\caption{Variation of total, upper and lower branch conductance with the unknown hopping parameter $\textnormal{J}_3$ for \textbf{(a)} system- dephasing B\"{u}ttiker bath coupling parameter $\gamma_P=0.5$.  Contour diagrams showing the positive and negative phases for the \textbf{(b)} Upper branch conductance and  \textbf{(c)} Lower branch conductance for the dephasing B\"{u}ttiker 
        bath.  The red region specifies the place where the conductance is  $<-10^{-4}$ and the blue region specifies conductance value $>10^{-4}$.
        \textbf{(d)} Variation of conductance for lossy channel coupling $\lambda=0.3$.
        \textbf{(e)} Phase diagram for the Upper branch conductance of the system with lossy channels .
			\textbf{(f)} Variation of the device efficiency in percentage  with the strength of environmental interaction $\gamma_P (\lambda).$ {The Yellowish (Off- white) part in the phase diagrams corresponds to currents with magnitude $<10
            ^{-4}$, which are considered too small to be measured properly in the experiment.}}
		\label{Environment_fig}
	\end{figure}

Here, the W matrix is given as,

\begin{align}
		W_{n,n}&=\gamma \gamma_P (|G^{R}_{n1}|^2+|G^{R}_{n4}|^2)+\gamma_P^2\sum_{m\ne n} |G^{R}_{nm}|^2 \nonumber \\
		W_{n,m}&=-\gamma^2_P  |G^{R}_{nm}|^2,  \forall n\ne m	
    \end{align}
    The results corresponding to this configuration are shown in Fig.~\ref{Environment_fig}. 
    
    In Fig.~\ref{Environment_fig} (a), it is observed that introducing dephasing via zero-current B\"{u}ttiker voltage probes shifts the point at which the upper and lower branch conductances change sign, a shift that directly impacts the device functionality since the detection of $\textnormal{J}_3$ relies on the current's sign reversal. Nevertheless, even for a relatively strong system–B\"{u}ttiker bath coupling of $\gamma_P = 0.5$, this shift remains modest. Also,the absolute magnitude  of the conductance reduces significantly in the presence of the external probes. Examining the phase diagrams in Fig.~\ref{Environment_fig} (b) and (c), which display the negative and positive values of the upper and lower branch conductances respectively, it is found that the zero crossings of both conductances begin to shift gradually as $\gamma_P$ is increased. This trend continues slowly until $\gamma_P \sim \gamma$, beyond which the zero crossings move more rapidly. Also, the shift is not symmetric, the divergence in the lower branch conductance occurs significantly faster than in the upper branch conductance. Additionally, if the coupling becomes very strong, the device eventually loses its functionality.

%%%%%%%%%%%%%%%%%%%%%%%%%%%%%%%%%%%%%%%%%%%%%%%%%%%%%%%%%%%%%%%%%%%%%%%%%%%%%%%%%%%%%%%%%%%%%%%%%%%%%%%%%%%%%%%%%%%%%%%%%%%%%%%%%%%%%%%%%%%%%%%%%%%%%%%%%%%%%%%%%%%%%%%%%%%%%%%%%%%%%%%%%%%%%%%%%%%%%%%%%%%%%%%%%%%%%%%%%%%%

	\subsection{Effect of lossy channels}
	To model a stronger environmental influence on our device, lossy channels \cite{Bijay_lossy_PhysRevB.110.235425} are introduced in the system, allowing particles to be lost within the device. Such a phenomenon can occur, for instance, in electron–phonon interactions, where inelastic scattering can change the energy of an electron, preventing it from being detected by the device \cite{datta_quantum_transport,Bijay_lossy_PhysRevB.110.235425}. To implement these particle leaks at all sites,  virtual leads similar to those defined in Eq.~\eqref{Bath hamiltonian} are utilized. However, in this case,  the zero-current constraint is not imposed on the virtual leads. Instead, the chemical potential of each virtual lead is set to the on-site chemical potential $\mu$, and the analysis is carried out under this condition. Consequently, the retarded Green's function is modified as follows:

	\begin{align}
		G^R=\left(E-H+\frac{i}{2}( \Gamma_1+ \Gamma_4)+\frac{i}{2}\Gamma_0 \right)^{-1}
	\end{align}
	 The correlation matrix in this case is given by  
\begin{equation}
    G^N = f_L G^R \Gamma_1 G^A + f_R G^R \Gamma_4 G^A + f_{\mathrm{eq}} G^R \Gamma_0 G^A ,
\end{equation}  
where $(\Gamma_{0})_{n,n} = \lambda$ for all $n$. Here, $\lambda$ quantifies the strength of the system’s coupling to the lossy leads, and $f_{eq}$ is the Fermi function corresponding to the onsite chemical potential $0$.
Due to particle losses, Kirchhoff’s rule does not necessarily hold, and  the conductances are redefined as averages:  
\begin{align}
    \mathcal{G}_T &= \frac{\mathcal{G}_1 + \mathcal{G}_4}{2},
    &&\mathcal{G}_U = \frac{\mathcal{G}_{1 \to 2} + \mathcal{G}_{2 \to 3} + \mathcal{G}_{3 \to 4}}{3}, 
    &&\mathcal{G}_D= \mathcal{G}_{1 \to 4}.
\end{align} 
	Figure~\ref{Environment_fig} (d) shows that the deflection of the zero-crossing points for the upper and lower branch conductances is much larger in the presence of lossy channels compared to the dephasing case, as expected since lossy channels represent stronger environmental influences. Additionally, the branch conductances no longer sum to give the total conductance, which is a natural consequence of allowing particle loss in the system. This effect is further highlighted in the phase diagram of Fig.~\ref{Environment_fig} (e), where even a small $\lambda$ causes the zero-crossing points to diverge from the ideal case of maximal device efficiency. The difference between the dephasing and lossy-channel scenarios is particularly evident when comparing Figs.~\ref{Environment_fig} (c) and \ref{Environment_fig} (e). Despite this significant environmental influence, the device continues to function effectively for relatively moderate virtual-bath couplings $\lambda$. To quantify this performance, we define a normalized efficiency parameter $\eta = (1 - \Delta \textnormal{J}_3 / \textnormal{J}_3^0)$, where $\Delta \textnormal{J}_3$ measures the difference between the two values of $\textnormal{J}_3$ at which the conductances of the upper and lower branches change sign, and  $\textnormal{J}_3^0$ is the ideal value which is $1$ for the given setup. So,
    for the ideal case, $\eta = 1$. As shown in Fig.~\ref{Environment_fig} (f), the device maintains good functionality (around 80\%) up to substantial environmental coupling, $\gamma_P,~ \lambda \sim \gamma$, indicating that it can reliably measure the unknown parameter even under strong environmental interference. Also, as expected, the efficiency is much higher in the dephasing case.
%%%%%%%%%%%%%%%%%%%%%%%%%%%%%%%%%%%%%%%%%%%%%%%%%%%%%%%%%%%%%%%%%%%%%%%%%%%%%%%%%%%%%%%%%%%%%%%%%%%%%%%%%%%%%%%%%%%%%%%%%%%%%%%%%%%%%%%%%%%%%%%%%%%%%%%%%%%%%%%%%%%%%%%%%%%%%%%%%%%%%%%%%%%%%%%%%%%%%%%%%%%%%%%%%%%%%%%%%%%%
	\section{Current Results at finite voltage and temperature}\label{finite_volt_temp_sec}
	\begin{figure}[b]
		\centering 
		\subfloat []
		{\includegraphics[width=0.32\linewidth,height=0.27\linewidth]{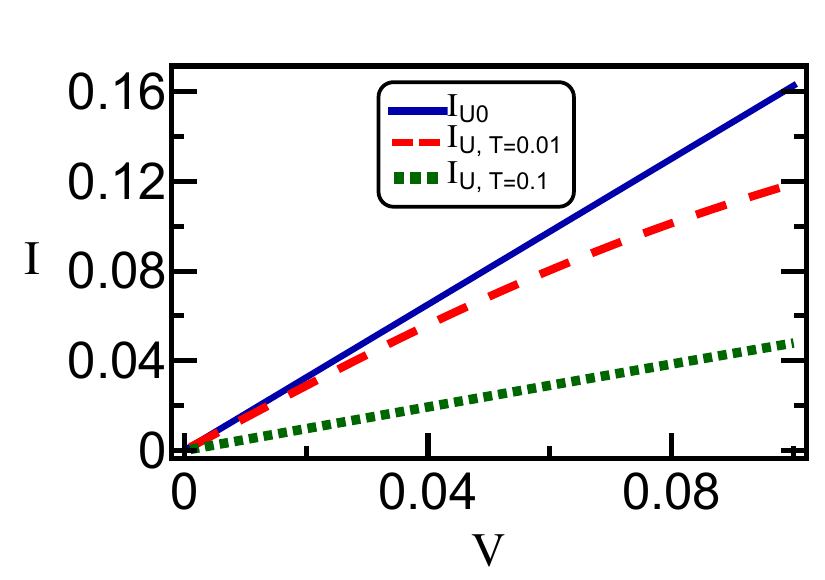}}
		\subfloat []
		{\includegraphics[width=0.32\linewidth,height=0.27\linewidth]{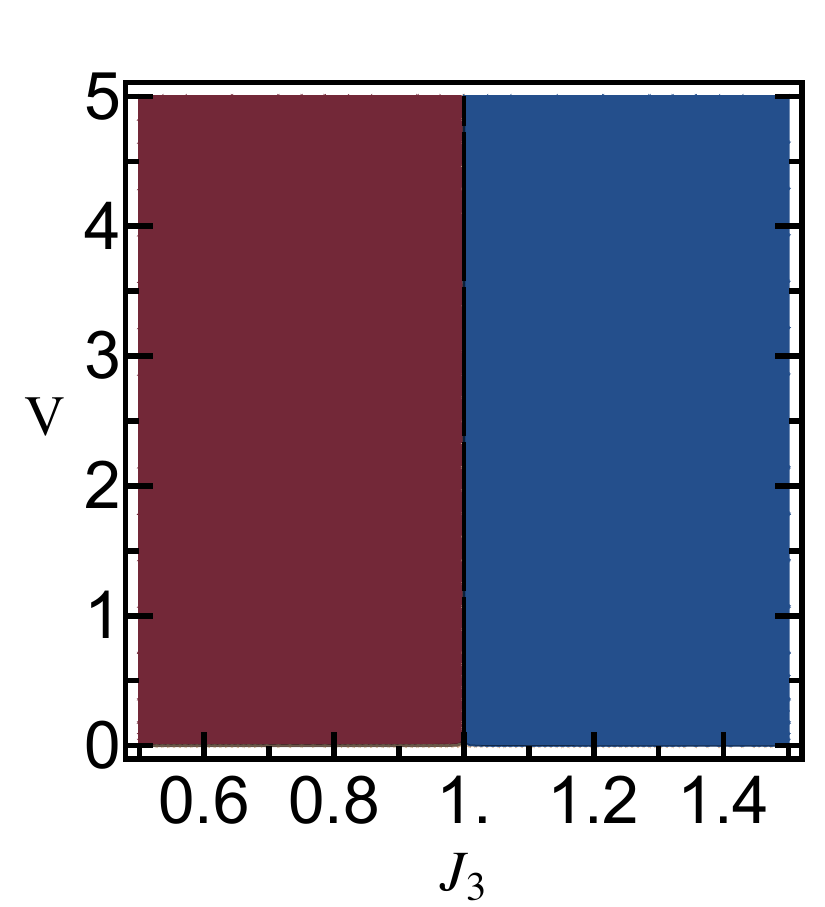}}
		\subfloat []
		{\includegraphics[width=0.32\linewidth,height=0.27\linewidth]{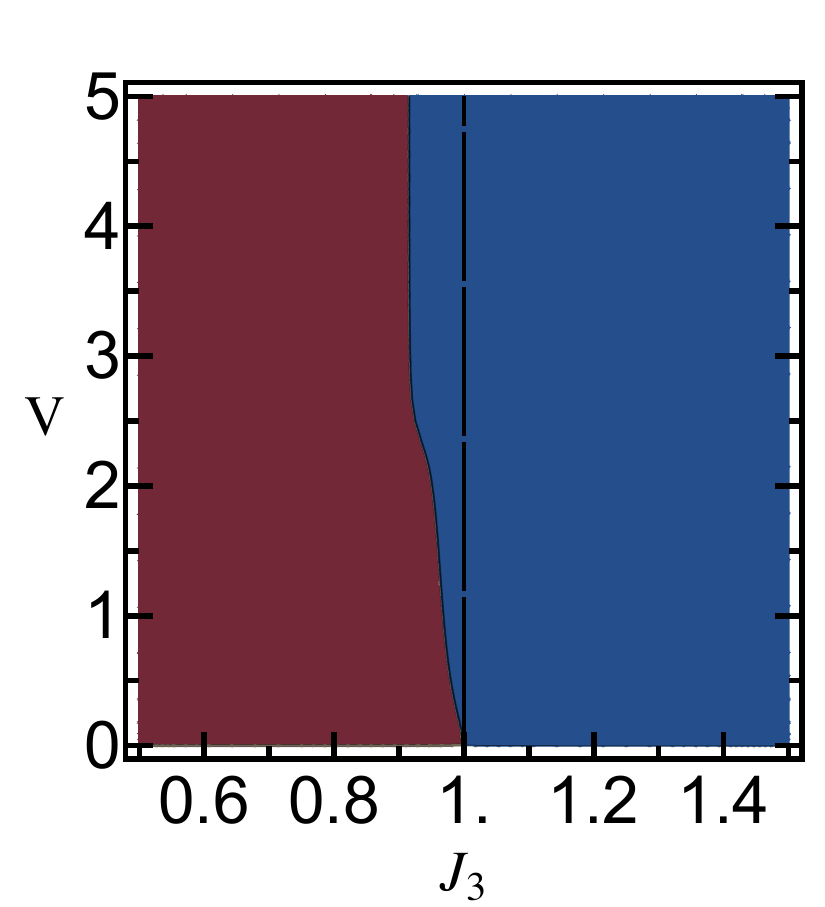}}
        \\
        \subfloat []
		{\includegraphics[width=0.32\linewidth,height=0.27\linewidth]{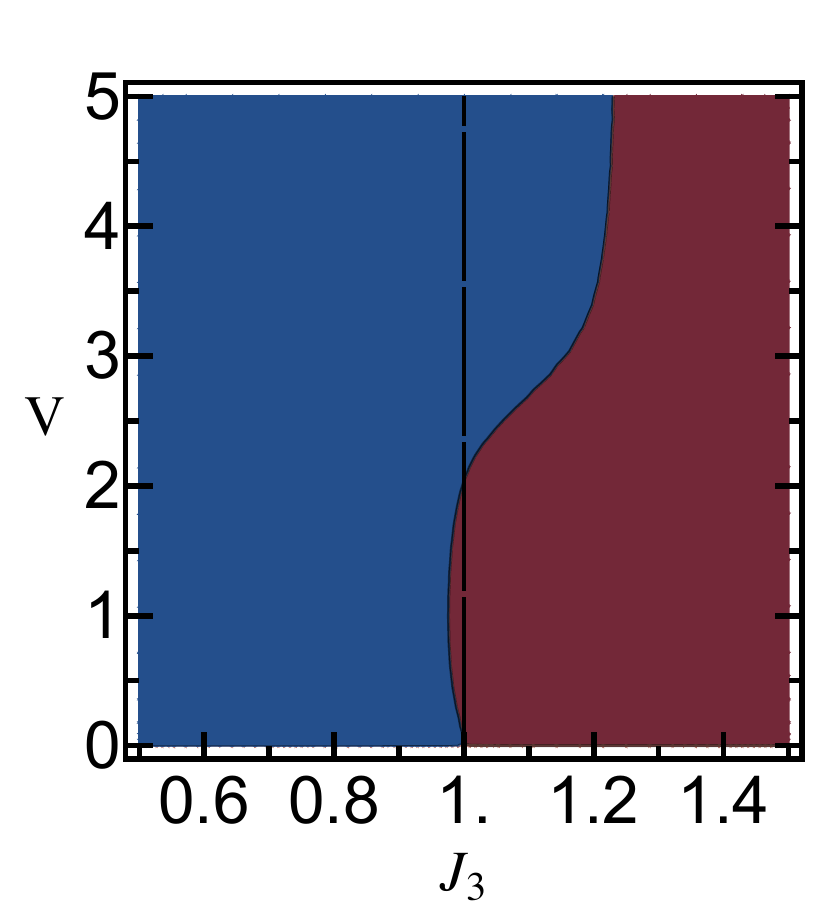}}
		\subfloat []
		{\includegraphics[width=0.32\linewidth,height=0.27\linewidth]{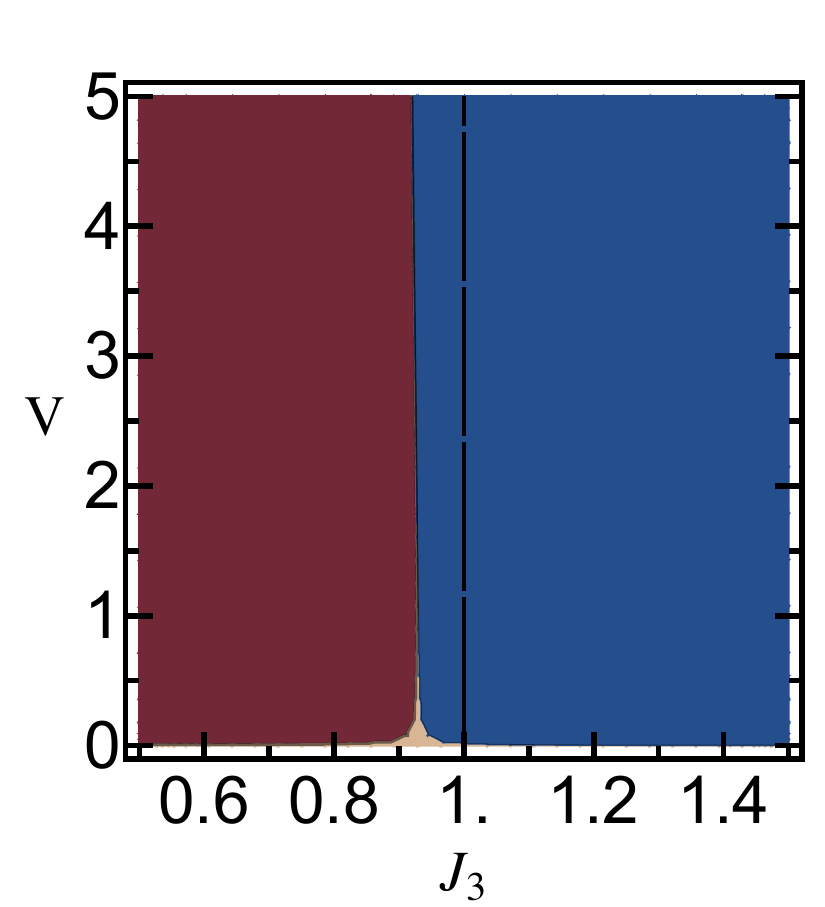}}
		\subfloat []
		{\includegraphics[width=0.32\linewidth,height=0.28\linewidth]{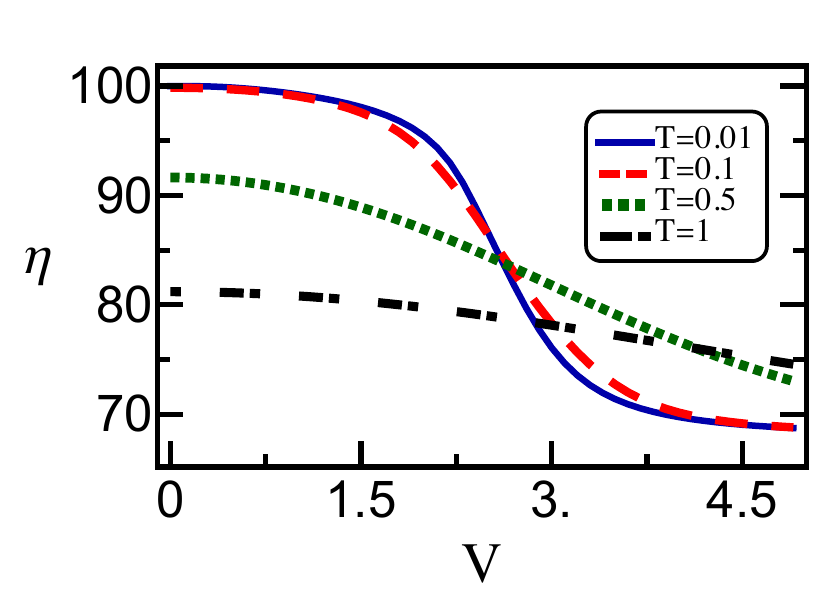}}
		\caption{\textbf{(a)} Variation of upper branch particle current with voltage for the ideal case and the numerical solutions.  Phases showing the negative and positive regions of the  upper branch current for the \textbf{(b)} ideal case, \textbf{(c)} numerically exact case. Phase diagram for \textbf{(d)} lower branch current and \textbf{(e)} upper branch current at higher temperature $T=1$. \textbf{(f)} Variation of device efficiency in percentage with voltage for different temperature values. If not specified otherwise, $T=0.01$. {The red region specifies the place where the current is  $<-10^{-4}$, the blue region specifies current value $>10^{-4}$, and the Yellowish (Off- white) part in the phase diagrams corresponds to currents with magnitude $<10
            ^{-4}$, which are considered too small to be measured properly in the experiment.}} 
		\label{voltage_and_temp}
	\end{figure}
	We now examine the performance of the device under arbitrary voltage bias and finite temperature. Operating in this regime typically introduces additional energy levels that contribute to the current, and as a result, the circulation arising primarily from the additional energy-degeneracy levels is challenged by parallel currents originating from other levels \cite{Upadhyay_2024}. We will discuss this in more detail in the next subsection. First, the following paragraph summarizes the results observed for finite voltage and temperature.  \par In Fig.~\ref{voltage_and_temp} (a), we compare the upper branch current obtained from the analytical formula derived in Eq.~\eqref{eqs_conductance} with the current calculated directly using the NEGF formalism.  Good agreement between the two results is observed in the low-bias, low-temperature regime. However, as either the voltage bias or temperature is increased, deviations appear, indicating that the device functionality may be affected under finite bias and temperature conditions. The phase diagrams for the upper and lower branch currents, shown in Figs.~\ref{voltage_and_temp} (c) and (d), respectively, further illustrate this effect, turning on the voltage bias shifts the zero-crossing points of both currents, with the shift being asymmetric and more pronounced in the lower branch. This behavior contrasts sharply with the ideal upper branch phase shown in Fig.~\ref{voltage_and_temp} (b), where the voltage has no impact on device functionality. Considering higher temperatures, Fig.~\ref{voltage_and_temp} (e) compares the upper branch current at $T = 1$ with that at $T = 0.01$, revealing that even at negligible bias, elevated temperatures significantly reduce device efficiency. To quantify these effects, Fig.~\ref{voltage_and_temp} (f) shows the variation of efficiency with voltage for different temperatures: at low bias, efficiency is nearly 100\% at low temperature but drops to around 80\% at high temperature. Interestingly, at higher voltages, the decline is steeper for the low-temperature case, eventually falling below the high-temperature efficiency for sufficiently large voltages. We now try to physically understand why the efficiency of the device reduces in the high voltage regime.
\subsection{Physical Insights into Voltage-Dependent Behavior and Environmental Effects}
\begin{figure}[b]
		\centering 
	\subfloat []{\includegraphics[width=0.32\linewidth,height=0.27\linewidth]{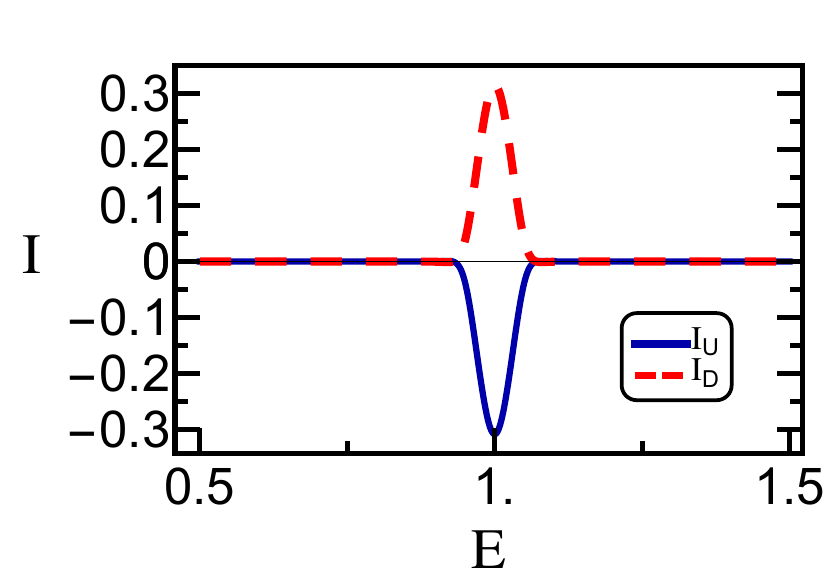}}
        \subfloat []
{\includegraphics[width=0.32\linewidth,height=0.27\linewidth]{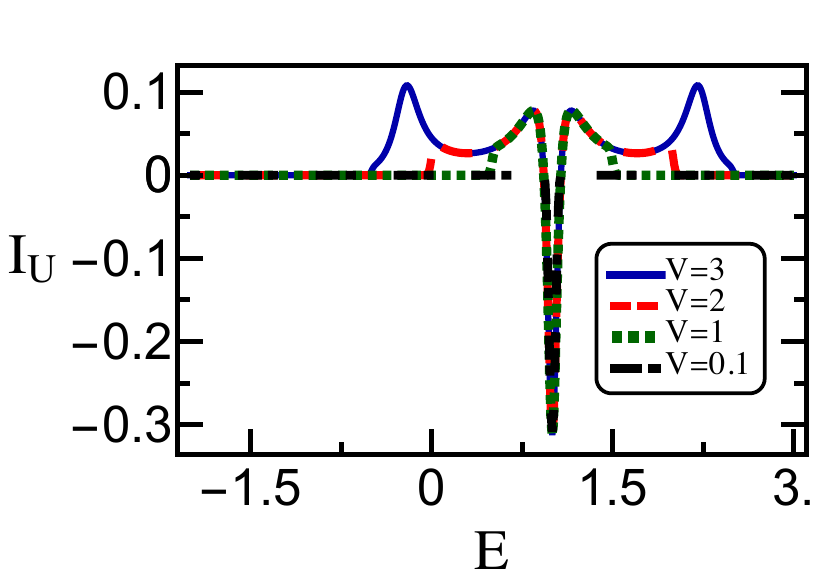}}        \subfloat []{\includegraphics[width=0.32\linewidth,height=0.27\linewidth]{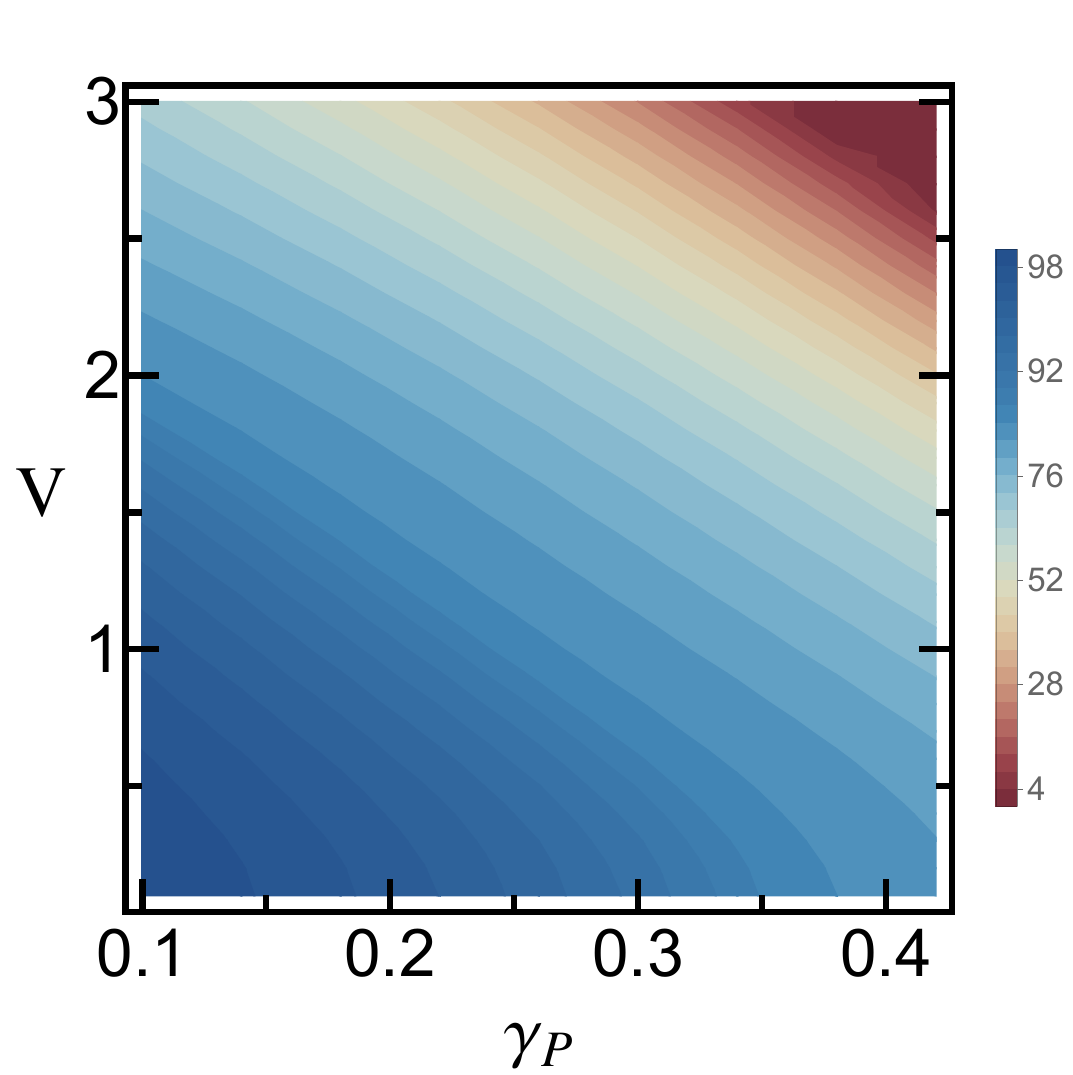}}\\
	    \subfloat []{\includegraphics[width=0.32\linewidth,height=0.27\linewidth]{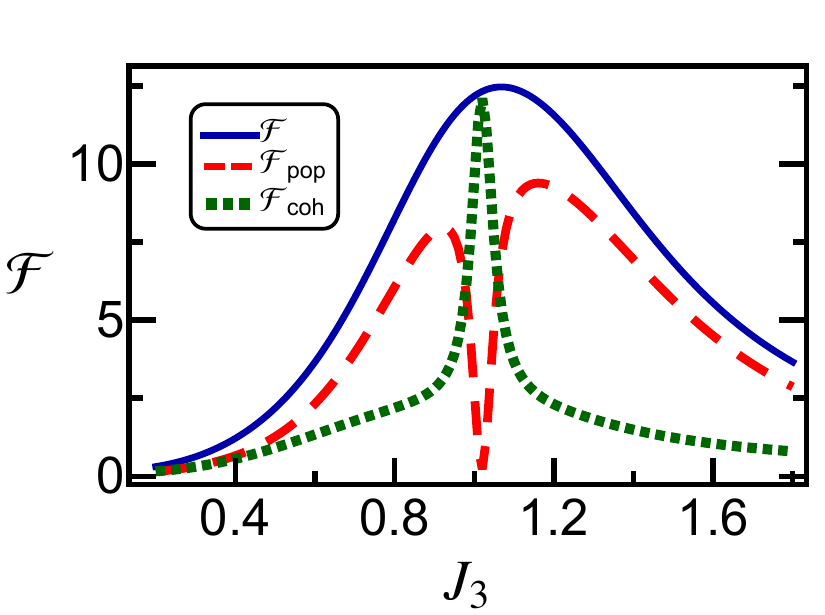}}
        \subfloat[]
        {\includegraphics[width=0.32\linewidth,height=0.27\linewidth]{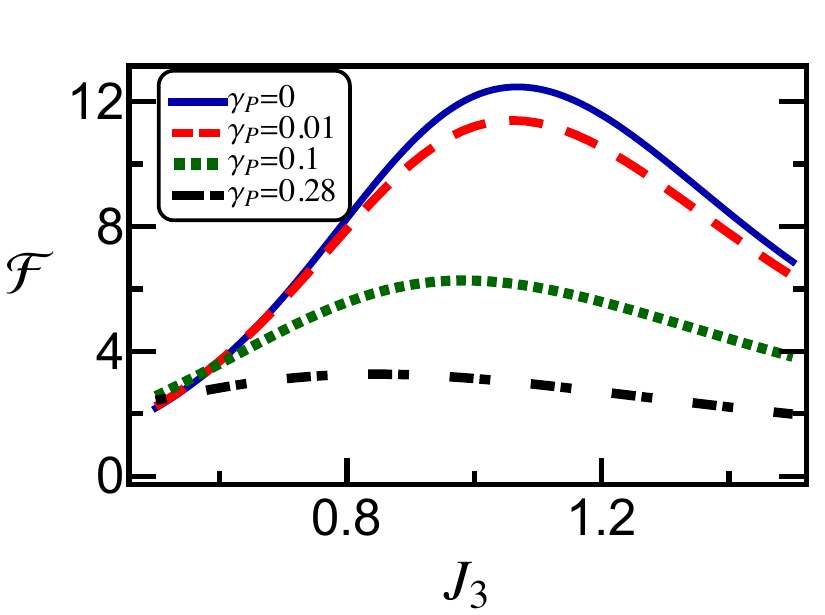}}
	 \subfloat []{\includegraphics[width=0.32\linewidth,height=0.27\linewidth]{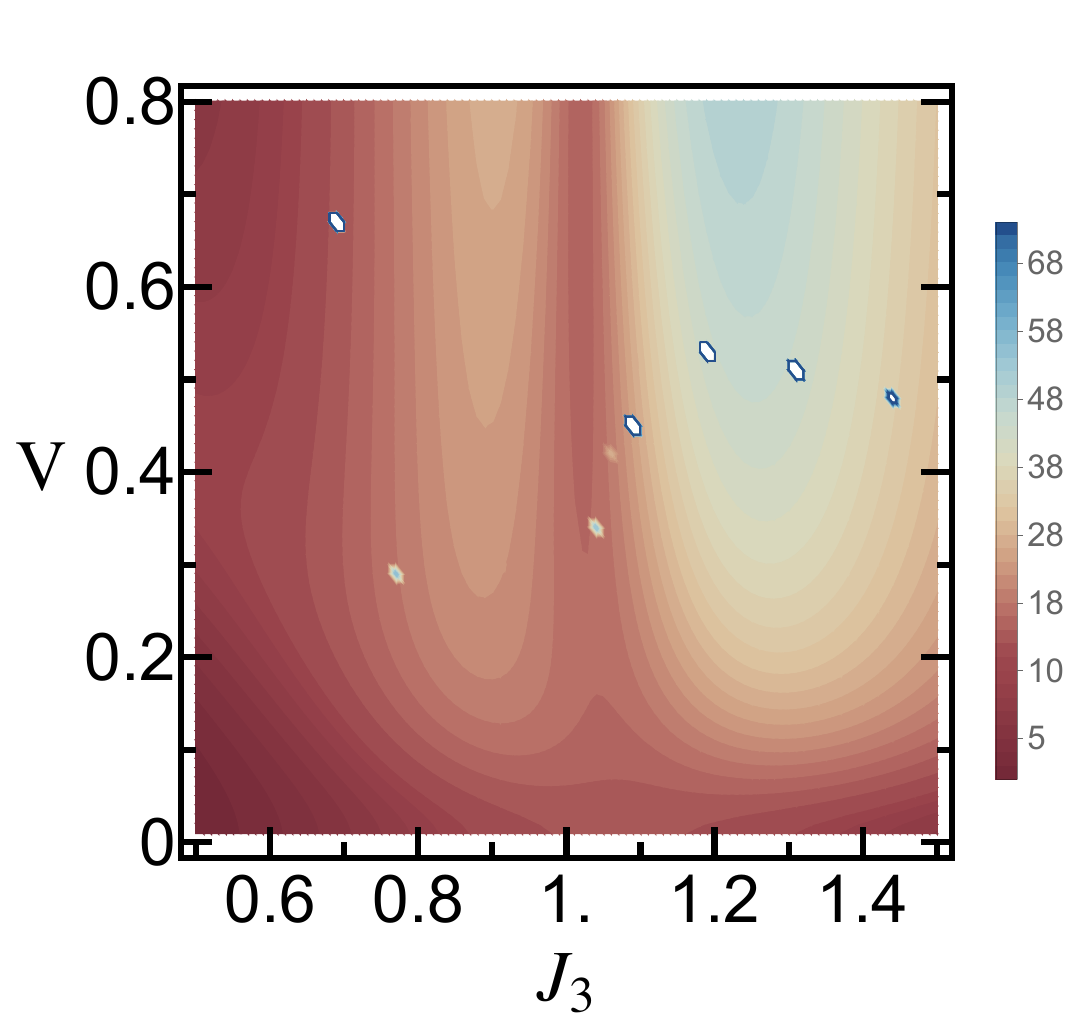}}
	 
	\caption{ \textbf{(a)} Energy dependent upper and lower branch current at V=0.1, $\textnormal{J}_3=0.98$. \textbf{(b)} Energy dependent upper  branch current for different voltages  at $\textnormal{J}_3=0.98$. 
    \textbf{(c)} Contour plot showing the value of efficiency with variation in dephasing and voltage for temperature T=0.1. 
    \textbf{(d)} Variation of different components of the quantum Fisher information (QFI) $\mathcal{F}$  with $\textnormal{J}_3$. 
    \textbf{(e)} Variation of total  QFI with $\textnormal{J}_3$ for various dephasing $\gamma_P$. \textbf{(f)} Contour plot showing the value of  total QFI with $\textnormal{J}_3$ Voltage. All the currents are scaled by a factor of $2\pi$. Unless otherwise specified, the parameters are $V=0.01,T=0.01$.
    }	\label{voltage_and_dephasing}
	\end{figure}
%%%%%%%%%%%%%%%%%%%%%%%%%%%%%%%%%%%%%%%%%%%%%%%%%%%%%%%%%%%%%%%%%%%%%%%%%%%%%%%%%%%%%%%%%%%%%%%%%%%%%%%%%%%%%%%%%%%%%%%%%%%%%%%%%%%%%%%%%%%%%%%%%%%%%%%%%%%%%%%%%%%%%%%%%%%%%%%%%%%%%%%%%%%%%%%%%%%%%%%%%%%%%%%%%%%%%%%%%%%%

To interpret the results presented above, we analyze the energy-dependent contributions to the upper and lower branch currents as a function of applied voltage. In other words, the behavior of integrands in Eq. ~\eqref{eq_u_cuurent} is examined with energy for different voltage values.
At low voltages, only energy levels near the on-site chemical potential $\mu$ contribute significantly to transport. As shown in Fig.~\ref{voltage_and_dephasing} (a), the upper branch current flows opposite to the bias. This behavior arises from the interplay between geometrical asymmetry and the AEDP \cite{Upadhyay_2024}. As the voltage increases, additional energy channels become active, favoring parallel rather than CC, which begins to compromise the device’s functionality, as seen in Fig. ~\ref{voltage_and_dephasing} (b).

With further voltage increase, the positive current contributions outweigh the negative ones, causing the total upper branch current to become positive. This explains the observed shift of the zero-crossing point away from the balanced Wheatstone condition. This also tells us that if the energy levels other than the ones causing the additional degeneracy are sufficiently far from the degeneracy point, any device based on AEDP will not loose functionality for moderate voltages.
Finally, examining the general case of finite voltage and environmental dephasing, illustrated in Fig.~\ref{voltage_and_dephasing} (c), the device initially maintains an efficiency above $90\%$ for moderate voltage and dephasing. However, for larger combined values of voltage and dephasing, the efficiency drops sharply, rendering the device ineffective. 

 To further support our idea, we examine how the Quantum Fisher Information (QFI) varies with different system parameters. QFI quantifies the precision in determining a given  parameter \cite{qf_deriv:doi:10.1142/S0219749909004839}. { Since our system has no intrinsic interaction in the Hamiltonian, the Non-equilibrium steady state (NESS) can be written as a number preserving Gaussian state, which in the diagonal basis is given as (see appendix \ref{appendix:qfi}),
 \begin{align}
    \tilde{\rho}_{S} = \frac{e^{-\sum_{k} \omega_k\, a_k^\dagger a_k}}{\mathcal{Z}},
\end{align}
where, $\mathcal{Z}=Tr[\tilde{\rho}_{S} ]$, `$\omega_k$' specifies the different modes in the NESS. For the above state, 
}
the total QFI can be written as \cite{qf_deriv:doi:10.1142/S0219749909004839} (see appendix \ref{appendix:qfi}), 
\begin{align}
\mathcal{F}
&=\mathcal{F}_{\text{pop}}+\mathcal{F}_{\text{coh}} \nonumber\\
&=\sum_l \dot{\omega}_l^{\,2}\big(\langle n_l\rangle -\langle n_l\rangle^2\big)
+2\sum_{p\neq i}
\frac{\big(\langle n_p\rangle -\langle n_i\rangle \big)^2}{\langle n_p\rangle +\langle n_i\rangle -2\langle n_i\rangle \langle n_p\rangle }\,
|S_{ip}|^2 .
\label{eq:QFI_main}
\end{align}
where,  $\langle n_l \rangle=\frac{1}{1+e^{\omega_l}}$, $\partial a_p^\dagger=\sum_i S_{ip} a_i^\dagger$. The first term represents the population-based sensitivity signifying how the occupation numbers change with the estimated parameter, while the second term captures the sensitivity arising from coherences through the off-diagonal matrix elements $S_{ip}$, which become important whenever the steady state has non-trivial quantum correlations. Since the aim is to determine $\text{J}_2$ in this study, the differentiations are taken with respect to it.

Plotting the QFI as a function of $\text{J}_3$ in Fig.~\ref{voltage_and_dephasing}(d),  it is observed that in the low-temperature, low-bias regime, the total QFI reaches a maximum near the transition point of the circulating current. This directly supports our claim that this region is optimal for estimating the unknown parameter $\text{J}_2$, since the QFI shows that the estimation uncertainty is minimal here. Interestingly, the population contribution to the QFI drops sharply close to the transition, whereas the coherence contribution rises steeply. A similar dip in the population  was previously observed in Ref.~\cite{Quantum_Wheatstone_Bridge,e21030228,Our_circulation}. {It primarily occurs due to significant population transfer between the near degenerate states as the system approaches AEDP, as the near degenerate eigenstates are strongly mixed near such places.  It can also lead to other interesting physics like population inversion and ergotropy increment near AEDPs \cite{e21030228,Our_circulation}}. In our regime of interest, however, the sharp rise of the coherence term dominates the fall in the population term, leading to an overall enhancement of the total QFI around the transition. This indicates that the working of our device is strongly dependent on the quantum nature of our system.

A similar qualitative trend appears on varying the dephasing strength in Fig.~\ref{voltage_and_dephasing}(e). For weak dephasing, the QFI again peaks near the transition point, but as the interaction strength increases the QFI curve progressively flattens, removing the advantage near the transition. This aligns with our physical expectation that dephasing reduces the efficiency of the device, although the QFI-based estimate suggests that this loss of efficiency may occur more rapidly than what our direct efficiency measure predicts.

Finally, in the contour plot of QFI as a function of voltage and $\text{J}_3$ (see Fig.~\ref{voltage_and_dephasing}(f)), it is observed that at low voltages the QFI peak remains sharply centered near the transition point. As the voltage increases and additional energy channels become active, the QFI peak broadens, and at sufficiently large voltages a dip develops near the transition with maxima shifted to either side. This suggests that the drop due to population contribution becomes the dominant contributor in this regime.
This behaviour qualitatively mirrors our physical predictions, though a QFI-based efficiency landscape naturally differs in quantitative detail from our operational definition of device efficiency.

In conclusion, the QFI analysis agrees qualitatively with our physical predictions and suggests that the low-temperature, low-bias regime with weak interactions is the optimal operating window for our device. { While, QFI provides an upper bound on the achievable precision, independent of the specific measurement protocol. 
The measurement protocol we propose is based on branch-resolved particle current measurements. By monitoring the signs of the upper and lower branch currents and identifying the value of the tunable parameter $\textnormal{J}_3$ at which the branch currents reverse direction, one can directly determine the unknown parameter $\textnormal{J}_2$ using the Wheatstone balance condition discussed in the manuscript.
Importantly, near the current-reversal point, the population contribution to the QFI is strongly suppressed while the coherence contribution becomes dominant. This is directly relevant for our protocol, since branch currents depend explicitly on off-diagonal correlations of the steady-state density matrix. As a result, the enhancement of QFI in this regime reflects an increased classical Fisher information associated with branch current-based measurements, rather than a purely abstract upper bound.
Therefore, although we do not explicitly optimize over all possible measurement schemes, the observed QFI enhancement near the transition point has a clear operational meaning and directly supports the sensitivity of the proposed current-based detection protocol.}

\section{Conclusion}\label{conclusion_sec}

We have investigated a minimal fermionic system designed to detect an unknown hopping rate between two sites by analyzing current circulation. The device leverages the principle of geometric asymmetry, which plays a crucial role in generating circulating currents within the system. By exploiting the connection between the additional energy degeneracy point (AEDP) and the direction of current circulation, we establish a sensitive mechanism for parameter estimation.

Our analysis begins in the low-temperature, low-bias regime, where the chemical potentials of the baths are aligned near the degenerate energy level. In this regime, a balanced Wheatstone bridge condition is identified, marked by the reversal of direction of the circulating currents. This reversal provides a direct and robust signature for determining the unknown hopping strength, without requiring measurement of the absolute magnitude of the current, a feature that may significantly simplify experimental implementation.
\par We further explore the impact of {random fixed parameter fluctuations}, environmental interactions, including dephasing and particle losses, which are inevitable in realistic setups. Our results indicate that the device maintains functionality under moderate to strong { random fluctuations as well as} environmental influences, demonstrating the robustness of the detection scheme. However, for extremely strong environmental perturbations, the CC is suppressed, and the device loses its ability to reliably detect the unknown parameter.
Beyond the idealized regime,  general operating conditions are also probed, without restrictions on voltage or temperature. Remarkably, the device continues to function effectively for moderate value of voltage and temperature, highlighting its practical applicability in realistic experimental scenarios \cite{exp_same_PhysRevX.7.031001,exp_same_2}.  Our results are further supported by the QFI analysis, which interestingly also shows that in the immediate vicinity of the AEDP, the coherence contribution to the QFI rises sharply, while the population contribution correspondingly drops.
This underscores the significance of geometric asymmetry as a design principle, which can be harnessed to enhance the sensitivity and robustness of quantum metrology devices. However, when the applied voltage becomes too large, contributions from other energy channels dominate over the AEDP channel, reducing the effectiveness of the device. 
Importantly, the concept presented here is generalizable. While this study focuses on a simple fermionic system, the principles of asymmetry-induced current circulation and the use of local current reversals can potentially be extended to spin systems, bosonic lattices, or more complex interacting systems, opening avenues for broader applications in quantum sensing and precision measurement \cite{review_quantum_sensing_RevModPhys.89.035002}. { Though temperature does not directly influence AEDPs in the Hamiltonian, it does influence them indirectly by affecting the broadening of the energy levels. As such, temperature determination by studying CC may be possible in a future study}.
In summary, our study demonstrates that geometric asymmetry, combined with CC, provides a powerful and experimentally accessible tool for parameter estimation in finite quantum systems, even in the presence of environmental disturbances. These insights can guide the design of robust metrological devices and inspire future explorations into symmetry-driven sensing protocols across diverse quantum platforms.

%%%%%%%%%%%%%%%%%%%%%%%%%%%%%%%%%%%%%%%%%%%%%%%%%%%%%%%%%%%%%%%%%%%%%%%%%%%%%%%%%%%%%%%%%%%%%%%%%%%%%%%%%%%%%%%%%%%%%%%%%%%%%%%%%%%%%%%%%%%%%%%%%%%%%%%%%%%%%%%%%%%%%%%%%%%%%%%%%%%%%%%%%%%%%%%%%%%%%%%%%%%%%%%%%%%%%%%%%%%%
		\section{Acknowledgement}
VU thanks Dr. Jagannath Sutradhar for insightful and patient discussions.

%%%%%%%%%%%%%%%%%%%%%%%%%%%%%%%%%%%%%%%%%%%%%%%%%%%%%%%%%%%%%%%%%%%%%%%%%%%%%%%%%%%%%%%%%%%%%%%%%%%%%%%%%%%%%%%%%%%%%%%%%%%%%%%%
\appendix
\section{Derivation of Expression for conductance with dephasing Buttiker probes}
We start by mathematically writing the condition that the net current exchanged with a virtual probe at energy `E' is zero. So the following equation is solved,
\begin{align}\label{first_append}
I_n(E)=\frac{1}{2\pi}\left(\mathcal{T}_{n1}(E)(f^P_n(E)-f_L(E))+\mathcal{T}_{n4}(f^P_n(E)-f_R(E))+\sum_{m\ne n} T_{nm}(f^P_n(E)-f^P_m(E)) \right)=0
\end{align} where the  transmission has two kind of terms, real leads to probe, $\mathcal{T}_{n1}=Tr[\Gamma^P_nG^R\Gamma_1 G^A]$, and probe to probe, $T_{mn}=Tr[\Gamma^P_mG^R\Gamma^P_n G^A]$. Here, $f_n(E)$ denotes the fermi function of probes and real leads both. The above equation can be written in a matrix form to solve for the fermi functions of the probes,
\begin{align}
    \sum_mW_{nm} f^P_m=R_n 
\end{align}
where, $R_n=\mathcal{T}_{n1}(E)f_L(E)+\mathcal{T}_{n4}(E)f_R(E)=\gamma\gamma_P |G^R_{n1}|^2f_L(E)+\gamma\gamma_P |G^R_{n4}|^2f_R(E)$, and the W matrix is given as,
\begin{align}
		W_{n,n}&=\gamma \gamma_P (|G^{R}_{n,1}|^2+|G^{R}_{n,4}|^2)+\gamma_P^2\sum_{m\ne n} |G^{R}_{n,m}|^2 \nonumber \\
		W_{n,m}&=-\gamma^2_P  |G^{R}_{n,m}|^2,  \forall n\ne m	
    \end{align}
   So, the fermi functions of the virtual probes by inverting the matrix $W$. Now in the low-bias low-temperature regime, the current equation \eqref{first_append} can be written in terms of voltage difference,
    \begin{align}
I_n(E)=\frac{1}{2\pi}\left(\mathcal{T}_{n1}(E)(V^P_n-V_L)+\mathcal{T}_{n4}(V^P_n-V_R)+\sum_{m\ne n} T_{nm}(V^P_n-V^P_m) \right)=0
\end{align}
Rewriting this as,
     \begin{align}
\mathcal{T}_{n1}(E)(V^P_n-V_R)+\mathcal{T}_{n4}(V^P_n-V_R)+\sum_{m\ne n} T_{nm}(V^P_n-V_R)-\sum_{m\ne n} T_{nm}(V^P_m-V_R)=\mathcal{T}_{n1}(E)(V_L-V_R) \end{align}
So,  voltage of each probe can be found,
\begin{align}
V^P_n-V_R=\gamma \gamma_P\sum_{m}W^{-1}_{nm}|G^R_{m1}|^2V
\end{align}
Finally, the current entering the right lead is,
   \begin{align}
   I_R&=I_T=\frac{1}{2\pi}\left(\mathcal{T}_{LR}(E)(V_L-V_R)+\sum_{m} \mathcal{T}_{m4}(V^P_m-V_R) \right), \nonumber \\
\mathcal{G}_T&=\frac{1}{2\pi}\left(\mathcal{T}_{LR}(E)+\frac{1}{V}\sum_{m} \mathcal{T}_{m4}(V^P_m-V^R) \right)=\frac{1}{2\pi}\left(\gamma^2 |G^R_{1,4}|^2+\gamma^2 \gamma^2_p\sum_{m,n} |G^R_{m,4}|^2W^{-1}_{m,n}|G^R_{n1}|^2 \right)
   \end{align}
   Similarly the branch conductances are obtained by the occupancy Green function $G^N$,
   \begin{equation}
    G^N(E) = f_L G^R \Gamma_1 G^A + f_R G^R \Gamma_4 G^A + \sum_m f_m G^R \Gamma^P_m G^A ,
\end{equation}  
Now,
\begin{align}
\mathcal{G}_U(E)=\frac{\textnormal{J}_1}{\pi} Im(G^N_{12}(E))=\frac{\textnormal{J}_1}{\pi}\left(\gamma Im(G^R_{1,1}G^A_{1,2})+\gamma\gamma^2_P\sum_{mn} Im(G^R_{1,m}G^A_{m,2})W^{-1}_{mn}|G^R_{n1}|^2\right)
\end{align}
And similarly  $\mathcal{G}_D$ can be calculated.
\section{Derivation of the Quantum Fisher Information for Gaussian steady state}\label{appendix:qfi}
We start this analysis by assuming a number preserving Gaussian steady state. This is consistent with all the analysis presented in the manuscript, as interaction are modeled phenomenologically through virtual Buttiker baths rather than considering them explicitly in the Hamiltonian. This state is given below,
\begin{align}
    \rho_{S}=\frac{e^{-\sum_{ij} c^\dagger_i M_{ij}c_j}}{\mathcal{Z}}
\end{align}
where, $\rho_S$ is the steady state density matrix, $M$ is the effective steady state Hamiltonian, and $\mathcal{Z}$ is the normalisation factor, with $\mathcal{Z}=Tr(e^{-\sum_{ij} c^\dagger_i Mc_j})$. We can express the above density matrix in its diagonal basis by diagonalizing the effective Hamiltonian $M$. This yields
\begin{align}
    \tilde{\rho}_{S} = \frac{e^{-\sum_{k} \omega_k\, a_k^\dagger a_k}}{\mathcal{Z}},
\end{align}
where $\omega_k$ are the eigenvalues of the matrix $M$, and $U$ is the unitary transformation that diagonalizes it. The fermionic operators in the two bases are related by
\begin{align}\label{operator_relation}
    a_k &= \sum_j U_{k j}^\dagger\, c_j, &
    a_k^\dagger &= \sum_j U_{j k}\, c_j^\dagger,
\end{align}
with
\begin{align}
    \mathrm{diag}(\omega_1, \omega_2, \ldots) = U^\dagger M U.
\end{align}
We start by the expression for QFI derived in the study \cite{qf_deriv:doi:10.1142/S0219749909004839}, the QFI for a density matrix defined as, $\rho=\sum_\alpha \rho_\alpha |\psi_\alpha\rangle \langle\psi_\alpha  |$ is given as,
\begin{align}
    \mathcal{F}_\theta=\sum_\alpha  \frac{(\partial_\theta \rho_\alpha)^2 }{\rho_\alpha}+2\sum_{\alpha\ne \beta}\frac{(\rho_\alpha-\rho_\beta)^2}{\rho_\alpha+\rho_\beta} |\langle\psi_\beta |\partial_\theta|\psi_\alpha\rangle|^2 =\mathcal{F}^{pop}_\theta+\mathcal{F}^{coh}_\theta
\end{align}
Now solving this term by term, first the population term,
\begin{align}
\mathcal{F}^{pop}_\theta=\sum_\alpha  \frac{(\partial_\theta \rho_\alpha)^2 }{\rho_\alpha}=\sum_\alpha \rho_\alpha(\partial_\theta ln(\rho_\alpha))^2=\langle(\partial_\theta ln(\rho))^2 \rangle
\end{align}
Now,
\begin{align}
    \partial_\theta ln(\rho_\alpha)=-\sum_l \dot{\omega}_l n_l^\alpha+\frac{\dot{\omega}_l}{1+e^{\omega_l}}
\end{align}
where $n_l^\alpha$ is the value of $n_l=\{0,1\}$ for state $\alpha$.
So,

\begin{align}
    (\partial_\theta ln(\rho_\alpha))^2=\sum_{l,k}\dot{\omega}_l\dot{\omega}_k\left(n^\alpha_l n^\alpha_k-\frac{n^\alpha_l}{1+e^{\omega_k}}-\frac{n^\alpha_k}{1+e^{\omega_l}}+\frac{1}{(1+e^{\omega_k})(1+e^{\omega_l})} \right)
\end{align}
Now taking the expectation and realising that $\langle n_l \rangle=\frac{1}{1+e^{\omega_l}} $, we get,
\begin{align}
\mathcal{F}^{pop}_\theta=\sum_l \dot{\omega_l}^2(\langle n_l\rangle -\langle n_l\rangle^2)
\end{align}
Now evaluating the coherence part,
\begin{align}
    \mathcal{F}^{coh}_\theta=2\sum_{\alpha\ne \beta}\frac{(\rho_\alpha-\rho_\beta)^2}{\rho_\alpha+\rho_\beta} |\langle\psi_\beta |\partial_\theta|\psi_\alpha\rangle|^2 
\end{align}
First, we evaluate the matrix term, $\langle\psi_\beta |\partial_\theta|\psi_\alpha\rangle$. Using the definition,
\begin{align}
a_k^\dagger &= \sum_j U_{j k} c^\dagger_j \nonumber \\
\partial a_k^\dagger&=\sum_j \dot{U}_{jk}c^\dagger_j=\sum_{ij}\dot{U}_{jk}U^\dagger_{ij}a_i^\dagger=\sum_i S_{ik} a_i^\dagger
\end{align}
where, $S=U^\dagger \dot{U}, S^\dagger=-S$. Now the vector,
\begin{align}
    |\psi_\alpha\rangle&=\Pi_l (a_l^\dagger)^{n_l^\alpha}|0\rangle \nonumber \\
    \partial_\theta |\psi_\alpha\rangle&=\sum_p \Pi_{l>p} (a_l^\dagger)^{n_l^\alpha}[\partial_\theta( a_p^\dagger)^{n_p^\alpha}]\Pi_{k<p} (a_k^\dagger)^{n_k^\alpha} \nonumber\\
    &=\sum_{p: n^\alpha_p=1}\Pi_{l>p}(a_l^\dagger)^{n_l^\alpha}\left(\sum_i S_{ip} a_i^\dagger \right) \Pi_{k<p} (a_k^\dagger)^{n_k^\alpha}
    =\sum_{p: n^\alpha_p=1,i: n^\alpha_i=0}S_{ip}|\psi_\alpha^{p\to i}\rangle
\end{align}
where, $|\psi_\alpha^{p\to i}\rangle$ means the state is the same as $\psi_\alpha$ only the occupied state at `p' is emptied and the unoccupied at `i' is filled. We are neglecting the reordering phase here as it will be squared later and just give 1. Taking the inner product now,
\begin{align}
\langle\psi_\beta |\partial_\theta|\psi_\alpha\rangle=\sum_{p: n^\alpha_p=1,i: n^\alpha_i=0}S_{ip} \delta_{\beta,\alpha(p\to i)}
\end{align}
Putting back in the QFI definition,
\begin{align}
\mathcal{F}^{coh}_\theta=2\sum_{\alpha\ne \beta}\sum_{p: n^\alpha_p=1,i: n^\alpha_i=0}\frac{(\rho_\alpha-\rho_\beta)^2}{\rho_\alpha+\rho_\beta} |S_{ip}|^2 \delta_{\beta,\alpha(p\to i)}
\end{align}
\begin{align}
\mathcal{F}^{coh}_\theta
&=2\sum_{\alpha\ne \beta}\sum_{p: n^\alpha_p=1,,i: n^\alpha_i=0}
\frac{(\rho_\alpha-\rho_\beta)^2}{\rho_\alpha+\rho_\beta}
|S_{ip}|^2\delta_{\beta,\alpha(p\to i)} \nonumber \\
&=2\sum_{\alpha}\sum_{p: n^\alpha_p=1,,i: n^\alpha_i=0}
\frac{\big(\rho_\alpha-\rho_{\alpha(p\to i)}\big)^2}
{\rho_\alpha+\rho_{\alpha(p\to i)}}|S_{ip}|^2 .
\end{align}

We now note that the eigenvalues are separable as
\begin{align}
\rho_\alpha=\Pi_l \frac{e^{-\omega_l n_l^\alpha}}{1+e^{-\omega_l}},
\end{align}
and for the configuration differing by one particle–hole substitution $(p\to i)$, the ratio of the two eigenvalues is
\begin{align}
\frac{\rho_{\alpha(p\to i)}}{\rho_\alpha}
=e^{-(\omega_i-\omega_p)},
\end{align}
Hence,
\begin{align}
\frac{(\rho_\alpha-\rho_{\alpha(p\to i)})^2}{\rho_\alpha+\rho_{\alpha(p\to i)}}
=\rho_\alpha
\frac{\big(1-e^{-(\omega_i-\omega_p)}\big)^2}
{1+e^{-(\omega_i-\omega_p)}} .
\end{align}
Putting back,
\begin{align}
\mathcal{F}^{coh}_\theta=2\sum_{\alpha}\sum_{p: n^\alpha_p=1,,i: n^\alpha_i=0}\rho_\alpha
\frac{\big(1-e^{-(\omega_i-\omega_p)}\big)^2}
{1+e^{-(\omega_i-\omega_p)}}|S_{ip}|^2 .
\end{align}
Now to overcome the restriction in the sum $n^\alpha_p=1,n^\alpha_i=0$ explicitly we mutliply by a factor $n^\alpha_p(1-n^\alpha_i)$, putting back,

\begin{align}
\mathcal{F}^{coh}_\theta&=2\sum_{\alpha}\sum_{p \ne i}\rho_\alpha(n^\alpha_p(1-n^\alpha_i))
\frac{\big(1-e^{-(\omega_i-\omega_p)}\big)^2}
{1+e^{-(\omega_i-\omega_p)}}|S_{ip}|^2 \nonumber \\&=2\sum_{p \ne i}
\frac{\big(1-e^{-(\omega_i-\omega_p)}\big)^2}
{1+e^{-(\omega_i-\omega_p)}}|S_{ip}|^2(\langle n_p\rangle-\langle n_i\rangle\langle n_p\rangle)\nonumber\\&=2\sum_{p\neq i}
\frac{\big(\langle n_p\rangle -\langle n_i\rangle \big)^2}{\langle n_p\rangle +\langle n_i\rangle -2\langle n_i\rangle \langle n_p\rangle }\,
|S_{ip}|^2 .
\end{align}
\subsection{Connection between the effective Hamiltonian and the correlation matrix}
We now establish the connection between the correlation matrix and the effective Hamiltonian $M$. In the diagonal basis the correlation matrix is defined as,
\begin{equation}
    A_{ij} \equiv Tr\!\big(\tilde{\rho}_S\, a_j^\dagger a_i\big).
\end{equation}
Since the steady-state density matrix $\tilde{\rho}_S$ is diagonal in this basis, 
the correlation matrix elements are obtained directly as
\begin{equation}
    A_{ij} = \delta_{ij}\,\frac{1}{1+e^{\omega_i}},
\end{equation}
Now writing the above correlation matrix in terms of original fermionic operators $c$,
\begin{equation}
    A_{ij} =\sum_{k,l} Tr\!\big(\tilde{\rho}_S\,  U_{ kj}\, c_k^\dagger  U_{i l}^\dagger\, c_l\big)=\sum_{k,l} Tr\!\big(\tilde{\rho}_S\, U_{ kj}\, c_k^\dagger  c_lU_{i l}^\dagger \, \big) =\sum_{k,l} \,U_{i l}^\dagger \hat{C}_{lk}U_{ kj}
\end{equation}
This means that $A=U^\dagger CU$, where $U$ is the same matrix that diagonalises the effective Hamiltonian M. So, QFI can be found just by diagonalising the correlation matrix and dont need the explicit steady state density matrix. 
%%%%%%%%%%%%%%%%%%%%%%%%%%%%%%%%%%%%%%%%%%%%%%%%%%%%%%%%%%%%%%%%%%%%%%%%%%%%%%%%%%%%%%%%%%%%
	\bibliography{biblo}
\end{document}